\documentclass[aps,prb, twocolumn, showpacs, superscriptaddress]{revtex4-1}
\usepackage[utf8]{inputenc}
\usepackage{graphicx}
\usepackage{pstricks}
\usepackage{array}
\usepackage{natbib}
\usepackage{amsmath}
\usepackage{pifont}
\usepackage{dsfont}
\usepackage{multirow}
\usepackage{amssymb}
\usepackage{wasysym}
\usepackage{hyperref}

\def\be{\begin{equation}}
\def\ee{\end{equation}}

\newcommand{\nd}[1]{ \mathrm{{\textbf{#1}}} }
\newcommand{\hnd}[1]{ \hat{\mathrm{{\textbf{#1}}}} }
\newcommand{\hid}[1]{ \tilde{\mathrm{{\textbf{#1}}}} }

\usepackage{listings}
\definecolor{darkblue}{rgb}{0,0,.6}
\definecolor{darkred}{rgb}{.6,0,0}
\definecolor{darkgreen}{rgb}{0,.6,0}
\definecolor{red}{rgb}{.98,0,0}

\def\ssmall{\fontsize{8pt}{2pt}\selectfont}
\begin{document}

\title{Quantum Monte-Carlo for correlated out-of-equilibrium nanoelectronics devices}

\newcommand{\spsmsA}{Univ. Grenoble Alpes, INAC-SPSMS, F-38000 Grenoble, France}
\newcommand{\spsmsB}{CEA, INAC-SPSMS, F-38000 Grenoble, France}
\newcommand{\spsmsC}{LPTMC, UMR 7600 CNRS, Universit\'e Pierre et Marie Curie, 75252 Paris, France}
\newcommand{\spsmsD}{IPhT, CEA, CNRS, URA 2306, 91191 Gif-sur-Yvette, France}

\author{Rosario E. V. Profumo}
\affiliation{\spsmsA}
\affiliation{\spsmsB}
\author{Christoph Groth}
\affiliation{\spsmsA}
\affiliation{\spsmsB}
\author{Laura Messio}
\affiliation{\spsmsD}
\affiliation{\spsmsC}
\author{Olivier Parcollet}
\affiliation{\spsmsD}
\author{Xavier Waintal}
\affiliation{\spsmsA}
\affiliation{\spsmsB}
\date{\today}

\begin{abstract}

We present a simple, general purpose, quantum Monte-Carlo algorithm for out-of-equilibrium interacting nanoelectronics systems.
It allows one to systematically compute the expansion of any physical observable (such as current or density) in powers of the electron-electron interaction coupling constant $U$. It is based on the out-of-equilibrium Keldysh Green's function formalism in real-time
and  corresponds to evaluating all the Feynman diagrams to a given order $U^n$ (up to $n=15$ in the present work).
A key idea is to explicitly sum over the Keldysh indices in order to enforce the unitarity of the time evolution. 
The coefficients of the expansion can easily be obtained for long time, stationary regimes, even at zero temperature.
We then illustrate our approach with an application to the Anderson model, an archetype interacting mesoscopic system. 
We recover various results of the literature such as the spin susceptibility or the "Kondo ridge" in the current-voltage characteristics. In this case, we found the Monte-Carlo free of the sign problem even at zero temperature, in the stationary regime and in absence of particle-hole symmetry. The main limitation of the method is the lack of convergence of the expansion in $U$ for large $U$, {\it i.e.} a mathematical property of the model rather than a limitation of the Monte-Carlo algorithm. Standard extrapolation methods of divergent series can be used to evaluate the series in the strong correlation regime.

\end{abstract}
\maketitle

The field of electronic correlations is largely dominated by 
applications to strongly correlated
materials such as high-$T_c$ superconductors or heavy fermions.
 As a result, the large effort made by the 
community to build new numerical techniques to address correlations 
aims chiefly at reaching strongly correlated regimes for systems whose 
one-body dynamics is rather simple (the archetype of these systems 
being the Hubbard model). There are, however, many situations where 
the correlations are either small or moderate, yet their interplay 
with one-body dynamics might be very interesting. Examples include, 
for instance, the zero-bias anomaly in disordered systems\cite{lee1985}, the Fermi-
edge singularity in a quantum dot\cite{matveev1992}, a Kondo impurity embedded in an 
electronic interferometer\cite{takada2014} and possibly the 0.7 anomaly in a quantum 
point contact\cite{thomas1996}. While for a few situations, e.g. zero-dimensional (Kondo effects) 
and one-dimensional (Luttinger liquids) systems there exist exact analytical and 
numerical techniques\cite{schollwock2005,bulla2008}, the vast majority of these problems remains elusive to theoretical approaches. The aim of this article is to design  a technique that 
could address moderate interactions for a large variety of out-of-equilibrium situations.

A natural route for dealing with electron-electron interactions is to compute
the expansion of physical quantities in powers of the interaction coupling
constant, denoted hereafter by $U$. This expansion is traditionally written in
terms of Feynman diagrams. One can then compute the first orders, or try
various resummation strategies that have been elaborated in order to choose the
relevant Feynman diagrams for a given problem.\cite{gukelberg2015}
From a numerical point of view, systematic expansions
in powers of $U$ have also been intensively studied. 
In this context, various diagrammatic Monte-Carlo have been developed and studied\cite{Prokofiev08polaron, DiagQMC, KozikEPL10, van2012NatPhys}, 
which aim at explicitly summing the series of Feynman diagrams numerically, for example for the self-energy.
Concerning quantum impurity models, there has been an intense activity in the recent years
in the development of new continuous (mostly imaginary) time quantum Monte-Carlo techniques, based on an expansion
in $U$ (or around the strong coupling limit).
These new algorithms are of huge practical 
value in solving the self-consistent impurity problems that arise from 
the dynamical mean-field theory of correlated bulk systems\cite{rubtsov2005,werner2006,gull2008,gull2011},
even though they still suffer from the sign problem.
They have been extended to the non-equilibrium case in a relatively 
straightforward way, simply adapting the Monte-Carlo method to the 
Keldysh formalism \cite{muhlbacher2008,schiro2009,schiro2010,werner2009,werner2010}.
However, these out-of-equilibrium versions suffer from a severe dynamical sign problem,
compared to their equilibrium counterparts, which has severely limited their usage in practice.
In particular, they can not reach the long-time steady-state limit in several regimes
of parameters. Moreover, the approach of Ref.~\onlinecite{werner2009,werner2010} has
only been shown to work with sufficient accuracy for an Anderson impurity with {\it particle-hole} symmetry, i.e. a very special point of the phase diagram. 
More recently, bold diagrammatic Monte-Carlo for impurity models have also been
extended to the Keldysh context and used in combination with the
master equation for the density matrix \cite{cohen2014greenPRL,cohen2014greenPRB} to reach longer time.
Finally, testing these approaches in large systems, even at moderate interaction, 
remains also an open question.

In this paper, we first present a simple, systematic and general Monte-Carlo method to compute the first 10 to 15
coefficients of the expansion of any physical observable for a nanoelectronic system, in an out-of-equilibrium situation.
The system can be a nanoscopic system connected to leads or a quantum impurity in a (possibly self-consistently determined) bath.
Our method can be applied in various non-equilibrium contexts, 
either at short time after a quench, or in the long-time steady-state.
In particular, it can easily reach the steady-state limit, even at zero temperature, as well as any intermediate time. It is also not limited to particle-hole symmetric case.
The software developed can be seen as an extension of the Kwant package\cite{groth2014} to tackle
electron-electron interactions, or of the Triqs package \cite{TRIQS} to deal with non-equilibrium situations.
Second, we discuss the issue of the summation of the perturbative series of the physical quantity, which is well-known to be a prominent topic in the quantum many-body problem.
We will show that in the out-of-equilibrium Anderson model, for the parameters studied here,  the current through the dot or the density on it
have a finite apparent radius of convergence as a function of $U$.
We will also show that by simply modifying the quadratic part of the action (i.e. playing with the so-called
$\alpha$ term in \cite{rubtsov2005}), one can significantly extend the radius of convergence, 
hence in practice compute for higher values of the interaction.
Finally, we show that extrapolation technique for divergent series, e.g. Lindel\"of method, can also 
significantly improve the range of applicability of our method.

In section \ref{summary}, we summarize our method and explain the main differences between our
work and previous ones. This short section is mostly for QMC experts and can be skipped for people
new to the field.  Section \ref{model} introduces our models and notations as well as the basic
many-body perturbation expression that forms our starting point. This expression relates the 
interacting observables (such as current or magnetization) to the non-interacting Green's function of the system. 
Section \ref{onebody} discusses how to obtain the latter one, a prerequisite to any QMC scheme. While this step is relatively straightforward for simple impurity problems (the vast majority of the systems considered so far), its generalization to non-trivial geometries requires some care, or can  become a computationally prohibitive task. 
Section \ref{brute} discusses a direct calculation of the first few orders of the interaction expansion by a brute force numerical integration.
The discussion of the structure of the functions to be integrated will lead to a key insight for the Monte-Carlo.
Section \ref{sec:qmc} describes our QMC algorithm. 
In Section \ref{sec:bare} we use the QMC algorithm to calculate the first
10-15 terms in the expansion in powers of the interaction strength of the
local charge on an Anderson impurity in equilibrium. We analyze the radius of convergence of
the series in presence/absence of a mean-field term in the
non-interacting Hamiltonian. 
In section \ref{sec:kondo}, we use the method in the ouf-of-equilibrium regime, to obtain some results associated with the Kondo effect.
The article ends with a discussion and various appendices that contain some proofs or technical details.  

\section{Summary of the approach}
\label{summary}
Let us briefly sketch our algorithm and its properties. Non-QMC experts can  skip this part, since its content will be detailed and explained in the next sections.
We start with a general Hamiltonian 
$\hnd{H}(t)=\hnd{H}_0(t)+U \hnd{H}_{\rm int}(t)$
where $\hnd{H}_0(t)$ is a non-interacting quadratic Hamiltonian of an 
infinite system (typically a nanoelectronic system connected to several electrodes)  and $\hnd{H}_{\rm int}(t)$ contains the interacting part
which is switched on at $t=0$. We aim at calculating the expansion of an observable $Q$
(say the current or the local occupation of an orbital) in powers of $U$: 
\be 
Q(U)=\sum_{n=0}^{+\infty} Q_n U^n
\ee 
The $Q_n$ are given by many-body perturbation theory in the Keldysh formalism in the form
of a multi-dimensional integral of a determinant, according to Wick's theorem:
\be 
Q_n= \sum_{{\cal C}_{n}} W({\cal C}_n)\det \nd{M}_{n}({\cal C}_n)
\ee
The sum $\sum_{{\cal C}_{n}}$ contains an $n$-dimensional integral over internal times $u_i \in [0,t]$ as well as a sum over $n$ Keldysh indices
$a_i \in \{0,1\}$ and a sum over the different interaction matrix elements. $W({\cal C}_n)$ contains interaction matrix elements. $\det \nd{M}_{n}({\cal C}_n)$ is the determinant of a matrix built up with the non-interacting Green's function of $\hnd{H}_0(t)$. Our algorithm works as follows:

(1) We compute directly the $Q_n$. In contrast, the imaginary-time or real-time\cite{werner2010,schiro2009, schiro2010}  
continuous-time algorithms sample the partition function of the problem $Z$. In the real-time Keldysh formalism however, the partition function is $Z=1$ by construction and as
we shall see, its sampling is not well suited for obtaining the $Q_n$.
Technically, the integrand $\det \nd{M}_{n}({\cal C}_n)$ is concentrated
around times $u_i$ which are close to $t$ while in the sampling of $Z$ the
$u_i$ are spread over the full interval $[0,t]$.
This first step ensures that our technique converges well as $t\rightarrow \infty$ and that this limit can be taken order by order in $U$.

(2) In the Keldysh formalism, one typically starts from a non interacting system, switches on the interaction, let the system evolve for a time $t$, 
measures the observable and then evolves back to the non-interacting initial state. The Keldysh indices $a_i$ are reminiscent of the two evolutions [from 0 to $t$ ($a_i=0$) and back to 0 ($a_i=1$)]. The time evolution is unitary. To keep this unitarity order by order, we choose to {\sl sum explicitly over the Keldysh indices}.
Hence our algorithm samples directly 
$|\sum_{\{a_i\}} W({\cal C}_n) \det \nd{M}_{n}({\cal C}_n)|$.
Indeed, performing the sum over the Keldysh indices only with the Monte-Carlo Markov chain implies that unitarity is only respected on average (i.e. not for a single configuration), and in other words relies on the Monte-Carlo
to perform massive cancellations.
Obviously the explicit sum comes at an exponential cost: for each Monte-Carlo move one needs
to calculate $2^n$ terms. However, we shall see that the gain in signal to noise ratio more than compensates for this additional computational cost. In the problems that we have computed so far, the dynamical sign/phase problem entirely disappears even for large times. From a
diagrammatic perspective, the summation over Keldysh indices also implements the automatic cancellation of disconnected Feynman diagrams.

(3) We compute the $Q_n$ coefficients individually rather than $Q(U)$. 
Indeed, once the $Q_n$ are obtained, we can analyze the convergence of the series for $Q(U)$,
which is a {\it separate mathematical problem} and has nothing to do with the Monte-Carlo 
or any other technique used to obtain the $Q_n$. 
Moreover, in continuous-time algorithms, the interaction very often takes the form of a density-density interaction 
$\hnd{H}_{\rm int}= (\hnd{n}_\uparrow-\alpha_\uparrow)(\hnd{n}_\downarrow -\alpha_\downarrow)$
and very special values of $\alpha_\sigma$ must be used for the computation not
to be plagued by the sign problem\cite{Shi2013,Naomi1997}. For instance in Ref.~\onlinecite{werner2010} the
algorithm fluctuates wildly away from $\alpha_\sigma= 1/2$. Here we find that
the convergence of the series for $Q(U)$ depends strongly on the value of
$\alpha_\sigma$. Note that this is a property of the perturbation series and
therefore independent of the procedure used to obtain the $Q_n$ coefficients. In
practice, the dependence of the radius of convergence on the $\alpha_\sigma$
parameters can be used to access larger values of the interaction: we add (to
$\hnd{H}_0$) and substract (to $\hnd{H}_{\rm int}$) an on-site potential to
our Hamiltonian such that the full Hamiltonian is unchanged but the series
acquires a larger radius of convergence. This step allows us to tackle systems
away from the particle-hole symmetry point. It is not linked to the QMC
technique {\it per se} but to the choice of the initial quadratic Hamiltonian
around which one performs the interaction expansion.

\section{Models and basic formalism}
\label{model}
\begin{figure}[h]
   \includegraphics[width=6cm]{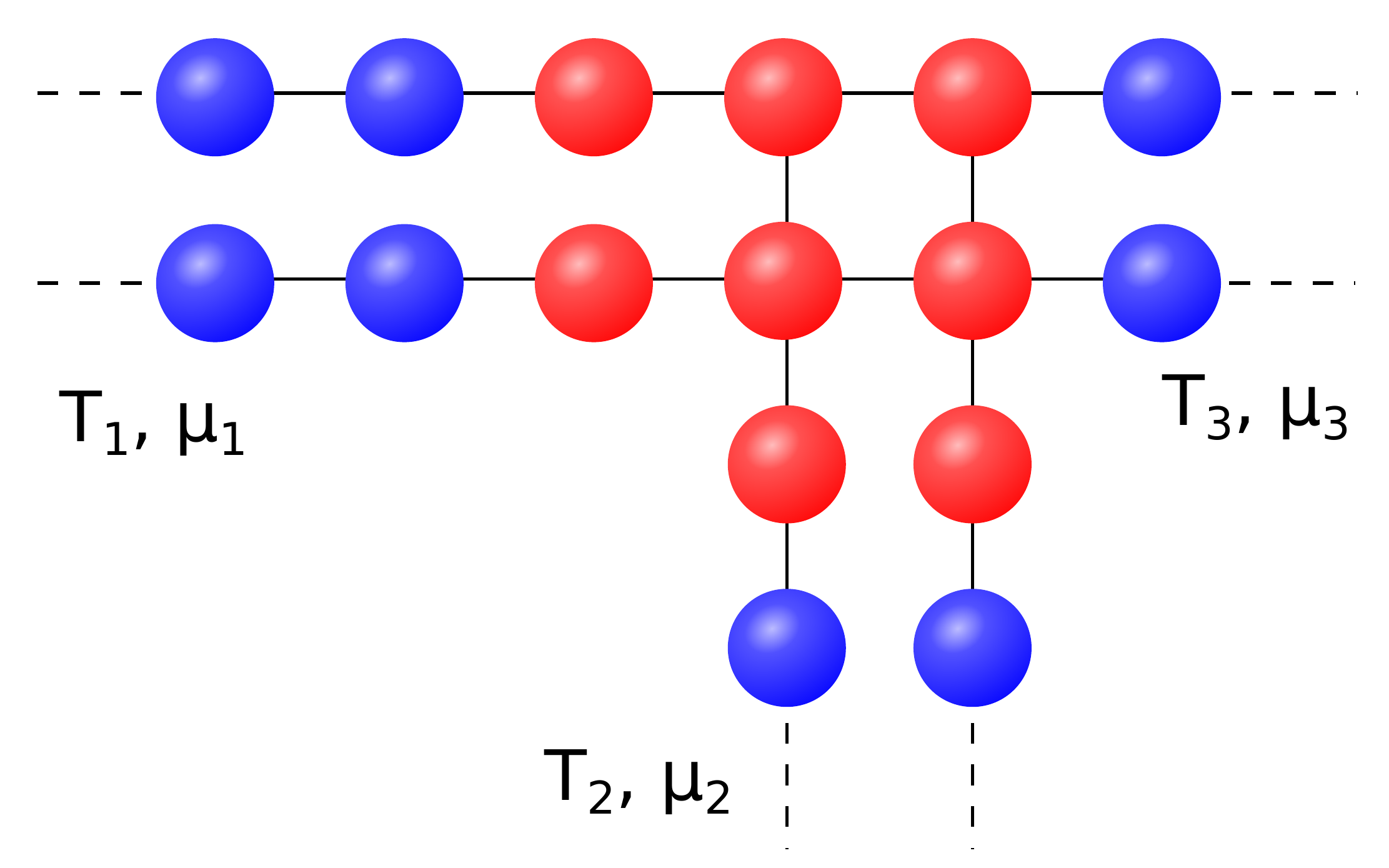}
\caption{
\label{fig:sys} Sketch of a typical mesoscopic system: a central interacting region (red) is connected
 to several (semi-infinite) non-interacting electrodes (blue) with finite temperatures $T_i$ and chemical potentials $\mu_i$.
 }
\end{figure}

\subsection{Models}

We consider a general time-dependent model for a confined nanoelectronic system connected to
metallic electrodes, following the approach of Ref.~\onlinecite{meir1992}. 
A sketch of a generic system is given in Fig.~\ref{fig:sys}.
The Hamiltonian
consists of a quadratic term and an electron-electron interacting term,
\begin{equation}
\hnd{H}(t)=\hnd{H}_0(t)+U \hnd{H}_{\rm int}(t)
\label{eq:H_general}
\end{equation}
where the parameter $U$ controls the magnitude of the interaction.
The non-interacting Hamiltonian takes the following form,
\be
\hnd{H}_0(t)=\sum_{i,j} \nd{H}^0_{ij}(t)\hnd{c}^\dagger_{i}\hnd{c}_{j}
\ee
where $c^{\dagger}_{i}$ ($c_{j}$) are the usual fermionic creation (annihilation)
operators of a one-particle state on the site $i$. The site index $i$ is general and can include different kinds of
degrees of freedom: space, spin, orbitals.
A crucial aspect is that the number of ``sites'' is {\it infinite} so that the non-interacting system has a well-defined
density of states (as opposed to a sum of delta functions for a finite system) while interactions only take place in a finite region. 
Typically, the system will consist of a  central part connected to semi-infinite periodic non-interacting leads. The dynamics of such non-interacting systems is well known and mature techniques exist to calculate both their stationary \cite{groth2014} and time-dependent properties
\cite{gaury2014}. The interaction Hamiltonian reads
\be
\hnd{H}_{\rm int}(t)=\sum_{ijkl} \nd{V}_{ijkl}(t)\hnd{c}^\dagger_{i}\hnd{c}^\dagger_{j}\hnd{c}_{k}\hnd{c}_l
\ee
In contrast to the non-interacting part, it is confined to a {\it finite} region. We also supposed that the interaction vanishes for negative time and is slowly or abruptly switched on at $t=0$. A typical system described by
Eq.~\eqref{eq:H_general} is a quantum dot where electrostatic gates confine the electrons in a small, highly interacting, region while the electrodes have high electronic density, hence weak interactions.

The techniques described below are rather general and will be discussed within the framework of Eq.~\eqref{eq:H_general}.
The practical calculations however will be performed on the following Anderson impurity models. Model A corresponds to one interacting site, ``0'' inside an infinite one-dimensional chain,
\begin{eqnarray}
\hnd{H}_{A}&=&\sum_{i=-\infty}^{+\infty}\sum_\sigma \gamma_i \hnd{c}^\dagger_{i,\sigma}\hnd{c}_{i+1,\sigma} 
+ h.c. + \epsilon_d (\hnd{n}_\uparrow+\hnd{n}_\downarrow) \nonumber \\
&+& U \theta(t) \hnd{n}_\uparrow \hnd{n}_\downarrow - h (\hnd{n}_\uparrow-\hnd{n}_\downarrow)
\end{eqnarray}
where 
\be
\hnd{n}_\sigma=\hnd{c}^\dagger_{0\sigma}\hnd{c}_{0\sigma}
\ee
$\epsilon_d$ is the level on-site energy, $h$ the (Zeeman) magnetic field and $\theta(t)$ is the Heaviside function so that the interaction is switched on at $t=0$. The hopping parameter $\gamma_i$ is equal to unity $\gamma_i=1$ for all sites except $\gamma_{-1}=\gamma_0=\gamma$. We apply a bias voltage $V_b$ between the two (Left and Right) electrodes which are characterized by their chemical potential $\mu_L=V_b$ and $\mu_R=0$ and temperature $T$. Model B is very close to model A with additional parameters $\alpha_\uparrow$,  $\alpha_\downarrow$,
\begin{eqnarray}
\hnd{H}_{B} 
&=& \sum_{i=-\infty}^{+\infty}\sum_\sigma \gamma_i \hnd{c}^\dagger_{i,\sigma}\hnd{c}_{i+1,\sigma}+h.c 
+\epsilon_d (\hnd{n}_\uparrow+\hnd{n}_\downarrow)
\nonumber \\
\label{eq:modelB}
&& -h (\hnd{n}_\uparrow-\hnd{n}_\downarrow)+U\theta(t)\left(\hnd{n}_\uparrow-\alpha_\uparrow\right)\left(\hnd{n}_\downarrow-\alpha_\downarrow\right)   
\end{eqnarray}
One easily realizes that the two models are in fact equivalent in the stationary limit for $\alpha_\uparrow =\alpha_\downarrow =\alpha$: $\hnd{H}_{B}(\epsilon_d,U,\alpha)=\hnd{H}_{A}(\epsilon_d-U\alpha,U)+U\alpha^2$. However they have very different large $U$ limit
at fixed (small) $\epsilon_d$: model A corresponds to the degeneracy point
between 0 and 1 electrons on the impurity (where Coulomb blockade is lifted)
while model B corresponds to the Kondo regime. More importantly, the
perturbation series in powers of $U$ of the same observable will be different between these two models, with different convergence radius for fixed $\epsilon_d$.
The $\alpha$ parameters have been introduced in Ref.~\onlinecite{rubtsov2005}, to improve the sign problem in imaginary-time Quantum Monte-Carlo.
An important energy scale for these models is the
(non-interacting) tunneling rate from the impurity to the reservoirs. It is
given by $\Gamma =\Gamma_L+ \Gamma_R$ with  $\Gamma_{L/R}=2\gamma^2
\sqrt{1-(\mu_{L/R}/2)^2}$.
 
\subsection{Interaction Expansion}
Our starting point for this work is a formal expansion of the out-of-equilibrium (Keldysh) Green's function in powers of electron-electron interactions. This is a standard step\cite{rammer1986} which we briefly sketch to introduce our notations.

Using the interaction representation, one defines $\hnd{c}_i(t)=\hnd{U}_0(0,t)\hnd{c}_i \hnd{U}_0(t,0)$ where $\hnd{U}_0(t',t)$ is the evolution operator from $t$ to $t'$ associated with $\hnd{H}_0$. Introducing the Keldysh index $a=0,1$, one defines the contour ordering for pairs $\bar t=(t,a)$: $(t,0)<(t',1)$ for all $t,t'$, $(t,0)<(t',0)$ if $t<t'$ and $(t,1)<(t',1)$ if $t>t'$. The contour ordering operator $T_c$ acts on products of fermionic operators $A,B,C\dots$ labeled by various ``contour times'' $\bar t_A=(t_A,a_A),\bar t_B,\bar t_C\dots$ and reorder them according to the contour ordering: $T_c(A(\bar t_A)B(\bar t_B)=AB$ if  $\bar t_A>\bar t_B$ and $T_c(A(\bar t_A)B(\bar t_B)=-BA$ if  $\bar t_A<\bar t_B$. The non-interacting contour Green's function is defined as
\be
g^c_{ij}(\bar t,\bar t')= -i \langle T_c \hnd{c}_i(\bar t) \hnd{c}^\dagger_j(\bar t') \rangle
\ee
where $\hnd{c}_i(\bar t)$ is just $\hnd{c}_i(t)$, the Keldysh index serving only to define the position of the operator after contour ordering. The contour Green's function has a matrix structure in $a,a'$ which reads
\be
g^c_{ij}(t,t')=
\left(
\begin{array}{cc}
g^T_{ij}(t,t') & g^<_{ij}(t,t') \\
g^>_{ij}(t,t')   &  g^{\bar T}_{ij}(t,t')
\end{array}
\right)
\ee
where $g^T_{ij}(t,t')$, $g^<_{ij}(t,t')$, $g^>_{ij}(t,t')$ and $g^{\bar T}_{ij}(t,t')$ are respectively the
time ordered, lesser, greater and anti-time ordered Green's functions. Efficient techniques to obtain these non-interacting objects for large systems will be discussed in the next section. Last, one defines the full Green's function 
$G^c_{ij}(\bar t,\bar t')$ with definitions identical to the above except that $\hnd{U}_0$ is replaced by $\hnd{U}$, the
evolution operator associated to the full Hamiltonian $\hnd{H}$. 
The fundamental expression for $G^c_{ij}(\bar t,\bar t')$ reads
\be
G^c_{ij}(\bar t,\bar t')= -i\langle T_c e^{ -i \int d\bar u\ U \hid{H}_{\rm int}(\bar u)} \hnd{c}_i(\bar t) \hnd{c}^\dagger_j(\bar t') \rangle
\ee
where the integral over $\bar u$ is taken along the Keldysh contour, i.e. increasing $u$ for $a=0$ and decreasing for $a=1$. $\hid{H}_{\rm int}(\bar u)$ is equal to $\hnd{H}_{\rm int}(u)$ with the operators $\hnd{c}_i,\hnd{c}^\dagger_j$ replaced by
$\hnd{c}_i(\bar u),\hnd{c}^\dagger_j(\bar u)$.  
\begin{widetext}
The expansion in powers of $U$ can now be performed,
\be
G^c_{ij}(\bar t,\bar t')= -i\sum_{n=0}^{+\infty} \frac{(-i)^n}{n!} U^n \sum_{\{a_i\}} (-1)^{\sum_i a_i}
\int du_1du_2\dots du_n \langle T_c \hid{H}_{\rm int}(\bar u_1) \hid{H}_{\rm int}(\bar u_2)\dots \hid{H}_{\rm int}(\bar u_n) \hnd{c}_i(\bar t) \hnd{c}^\dagger_j(\bar t') \rangle
\ee
Of particular interest to us are one-particle observables (say current or electronic density) which can be directly expressed in terms of the lesser Green's function at equal times:
\be
\label{eq:O}
O_{ij}\equiv \langle \hnd{U}(0,t) \hnd{c}^\dagger_i\hnd{c}_j \hnd{U}(t,0) \rangle=-i G^<_{ji}(t,t)
\ee
Note at this stage that the following derivation is presented for the one particle correlator, but can be straightforwardly generalized
to higher correlators.

To proceed, one evaluates the average of the (large) products of creation/destruction operators using Wick theorem,
in a form of the determinant of a $(2n+1) \times (2n+1)$ matrix,
\be
G^c_{ij}(\bar t,\bar t')= \sum_{n=0}^{+\infty} \frac{i^n}{n!} U^n \sum_{\{a_i\}} (-1)^{\sum_i a_i}
\int du_1du_2\dots du_n
\sum_{i_1j_1k_1l_1}V_{i_1j_1k_1l_1}(u_1)\dots \sum_{i_nj_nk_nl_n}  V_{i_nj_nk_nl_n}(u_n)
\det  \nd{M}_n
\label{eq:basic}
\ee
where $\nd{M}_n$ is given by
\be
\label{eq:M}
\nd{M}_{n} =\left(
\begin{array}{lllll} 
g^<_{k_1i_1}(\bar u_1,\bar u_1) & g^<_{k_1j_1}(\bar u_1,\bar u_1)  & g^c_{k_1i_2}(\bar u_1,\bar u_2) &...&g^c_{k_1 j}(\bar u_1,\bar t')  \\
g^<_{l_1i_1}(\bar u_1,\bar u_1) & g^<_{l_1j_1}(\bar u_1,\bar u_1)  & g^c_{l_1i_2}(\bar u_1,\bar u_2) &...&g^c_{l_1 j}(\bar u_1,\bar t') \\
g^c_{k_2i_1}(\bar u_2,\bar u_1) & g^c_{k_2j_1}(\bar u_2,\bar u_1)  & g^<_{k_2i_2}(\bar u_2,\bar u_2) &...&g^c_{k_2 j}(\bar u_2,\bar t')  \\
...&...&...&...&...\\
g^c_{k_ni_1}(\bar u_n,\bar u_1) & g^c_{k_nj_1}(\bar u_n,\bar u_1)  & g^c_{k_ni_2}(\bar u_n,\bar u_2) &...&g^c_{k_n j}(\bar u_n,\bar t')  \\
g^c_{l_ni_1}(\bar u_n,\bar u_1) & g^c_{l_nj_1}(\bar u_n,\bar u_1)  & g^c_{l_ni_2}(\bar u_n,\bar u_2) &...&g^c_{l_n j}(\bar u_n,\bar t')  \\
g^c_{i i_1}(\bar t,\bar u_1) & g^c_{i j_1}(\bar t,\bar u_1)  & g^c_{i i_2}(\bar t,\bar u_2) &...&g^c_{i j}(\bar t,\bar t')  \\
\end{array}
\right)
\ee 
and the zeroth order term is $g^c_{ij}(\bar t,\bar t')$. 
\end{widetext}
Eq.~\eqref{eq:basic} might look cumbersome at first sight, yet it is a compact expression: provided one knows how to calculate non-interacting Green's functions (which will be taken care of in the next section), Eq.~\eqref{eq:basic} expresses the full interacting Keldysh Green's function, hence the physical observables, in terms of integrals and sums of determinant of known quantities. All that remains is to find a suitable numerical way to perform those integrals and sums. For a local interaction present on $L$ sites, calculating all contributions to order $U^n$ in the stationary regime corresponds to a numerical complexity of the order of $L^n t^n$ where the measurement time $t$ has to be large enough for the effect of the electron-electron interaction to be well established. This can be performed using standard integration routines for the first 
few orders (In section \ref{brute} we calculate contributions $n=0,1,2,3,4$ for model A) but becomes quickly prohibitive for larger values of $n$. For larger orders, a stochastic sampling of the integrals is compulsory.

To simplify the notations, we introduce the notion of configuration ${\cal C}_n$
\be
{\cal C}_n=(i_1,j_1,k_1,l_1,u_1,\dots,i_n,j_n,k_n,l_n,u_n)
\ee
and note
\be
\sum_{{\cal C}_n} = \int_{0<u_1<\dots <u_n<{\rm max}(t,t')}
\kern -50pt
du_1du_2\dots du_n
\sum_{i_1j_1k_1l_1}\dots \sum_{i_nj_nk_nl_n}
\ee
Introducing,
\be
V({\cal C}_n)=\prod_{p=1}^n   V_{i_pj_p,k_pl_p}
\ee
we get the following compact expression:
\begin{eqnarray}
G^c_{ij}(\bar t,\bar t')= \sum_{n=0}^{+\infty} i^n U^n\sum_{\{a_i\}} (-1)^{\sum_i a_i} 
\times \nonumber
\\
\sum_{{\cal C}_{n}} V({\cal C}_n)\det \nd{M}_{n}({\cal C}_n,\{a_i\})
\end{eqnarray}
where the $n!$ factor has dropped out due to the ordering of the $u_i$.
Note that in the Keldysh formalism the partition function is unity which translates into
\be
0= \sum_{n=1}^{+\infty} i^n U^n\sum_{\{a_i\}} (-1)^{\sum_i a_i} \sum_{{\cal C}_{n}} V({\cal C}_n)\det \nd{P}_{n}({\cal C}_n,\{a_i\})
\ee
where the $2n\times 2n$ matrix $\nd{P}_{n}$ is identical to $\nd{M}_{n}$ with the last row and column deleted.
Actually, a much stronger statement can be made on $\nd{P}_{n}$: for any $n>0$ and configuration ${\cal C}_n$, one has,
\be
\label{eq:zero}
\sum_{\{a_i\}} (-1)^{\sum_i a_i} \det  \nd{P}_{n}({\cal C}_n,\{a_i\})=0
\ee
The proof is straightforward and standard: one first locates the largest time in the configuration ${\cal C}_n$, say $u_n$.
When $a_n$ goes from $0$ to $1$, the ordering of $\bar u_n$ with respect to the other times is unchanged
($\bar u_n$ is larger than all the times on the upper part of the contour and smaller than all those on the 
lower part of the contour), hence the contour Green's functions are unchanged and the matrix $\nd{P}_{n}$ is also  
unchanged. As a result of the $(-1)^{a_n}$ sign these two contributions cancel each other.

\section{The non-interacting Green's function}
\label{onebody}
In order to proceed with evaluating the interaction corrections to observables, the first step is
an efficient way to calculate the various real-time non-interacting Green's functions of the problem.
For a small dot problem or a DMFT model, this step is easy. For larger systems, this question is more delicate,
and it has been studied extensively\cite{gaury2014} and we briefly summarize the main aspects here.
Note that in Ref.~\onlinecite{gaury2014}, only quantum transport was of interest so that contributions coming from bound states could
have been omitted. Here however, they will have to be taken into account properly. 
\subsection{General method}
Our starting point for calculating non-interacting Green's functions is an expression that relates them
to the (Scattering) wave functions in the system \cite{gaury2014},
\begin{align}
&g_{ij}^<(t,t') =i\sum_\alpha \int \frac{dE}{2\pi}\ f_\alpha(E)  \Psi_{\alpha E}(t,i) \Psi^*_{\alpha E}(t',j)
\nonumber \\
&+i\sum_n f(E_n) \Psi_n(t,i)\Psi_n^*(t',j)
\label{eq:psi-less}
\end{align}
Here, $\alpha$ labels the various propagating channels of the leads, $\Psi_{\alpha E}(t,i)$ the scattering state at energy $E$ (in the electrode) and $f_\alpha(E)$ the corresponding Fermi distribution function.
$n$ labels a bound state of energy $E_n$ and wave functions $\Psi_n$.
The greater Green's function $g_{ij}^>(t,t')$ is obtained with an identical expression with the Fermi functions 
$f(E)$ replaced by $f(E)-1$. Efficient techniques for calculating the scattering wave functions $\Psi_{\alpha E}(t,i)$
have been designed so that these objects can be obtained for large systems ( $10^5$ sites\cite{gaury2014b}).
The bound states contribution was not considered in Ref.~\onlinecite{gaury2014} and will be discussed below.
The actual calculations performed in this article were restricted to a stationary non-interacting system, where
the above expression further simplifies into
\begin{align}
&g_{ij}^<(t-t') =i\sum_\alpha \int \frac{dE}{2\pi}\ f_\alpha(E)  \Psi_{\alpha E}(i) \Psi^*_{\alpha E}(j)e^{-iE(t-t')}
\nonumber\\
&+i\sum_n f(E_n) \Psi_n(i)\Psi_n^*(j)e^{-iE_n(t-t')}
\label{eq:psi-less-st}
\end{align}
Here again, the stationary wave functions $\Psi_{\alpha E}(i)$ are standard objects. They are in fact direct outputs of the Kwant software \cite{groth2014} which we use for their calculations. Once the lesser and greater Green's functions
are known, one completes the $2\times 2$ Keldysh matrix with the standard relations
\begin{align}
&g^T_{ij}(t,t')= \theta(t-t')g^>_{ij}(t,t') +
\theta(t'-t)g^<_{ij}(t,t')\\
&g^{\bar T}_{ij}(t,t')= \theta(t'-t)g^>_{ij}(t,t') +
\theta(t-t')g^<_{ij}(t,t')
\end{align}
To obtain those Green's function numerically, i.e. for many values of $t-t'$, one needs to perform the integration over the energy $E$ many times. 
In practice, the stationary wave functions are calculated once using Kwant and cached. The integration itself is performed using standard numerical routines. For the single site model A or B, the above technique in its full generality can be avoided: one can simply compute the Green's function in energy analytically and perform a numerical Fourier transform. We
have checked explicitely that both techniques provide identical non-interacting Green's functions in this special case.

\subsection{Bound states contribution}
The presence of the electrodes in the system is very important physically: it provides the system with a relaxation mechanism. Mathematically, the integral in Eq.~\eqref{eq:psi-less-st} mixes nearby energies so that the resulting
non-interacting Green's functions decay (and oscillate) at large times. However, in presence of a large enough confining energy (far from zero $\epsilon_d$ parameter in model A), true bound states can appear in the system. They have energies outside
of the electrode bands and therefore cannot hybridize with the plane waves of the electrodes. They satisfy
the stationary Schrodinger equation $\nd{H}^0 \Psi_n = E_n \Psi_n$ for the infinite system. Upon integrating
over the electrode degrees of freedom, they satisfy a simpler (yet non-linear) equation for the interacting region only:
\be 
\nd{H}^0 \Psi_n + \Sigma  (E_n) \Psi_n = E_n \Psi_n
\label{eq:bs}
\ee
where $\Sigma (E)$ is the retarded self-energy due to the electrode.
For a practical calculation, we do as follows: first we truncate $\nd{H}^0$ and keep the interacting region plus a rather
large (yet finite) fraction of the electrodes. We diagonalize the corresponding finite matrix and locate the eigenvalues
that are outside the conducting bands of the electrodes. These eigenvalues are used as initial guess and we compute the
bound states by iteratively solving Eq.~\eqref{eq:bs} until convergence. Note that there is an easy check to make sure that
one uses a complete basis of the problem: one must have,
\be
\sum_\alpha \int \frac{dE}{2\pi}\   |\Psi_{\alpha E}(i)|^2 + \sum_n  |\Psi_n(i)|^2=1
\ee
which is not verified if some bound states are forgotten. Note also that in most of this article, we focus on situations where there are no bound states in the system. This can be easily achieved by using leads which have a larger bandwidth than that of the central system, so that any bound state that could take place there hybridize with the continuum of the lead. 

\section{Analysis of the first terms of the perturbative series}
\label{brute}
\begin{figure}[h]
    \centering
    \includegraphics[width=8cm]{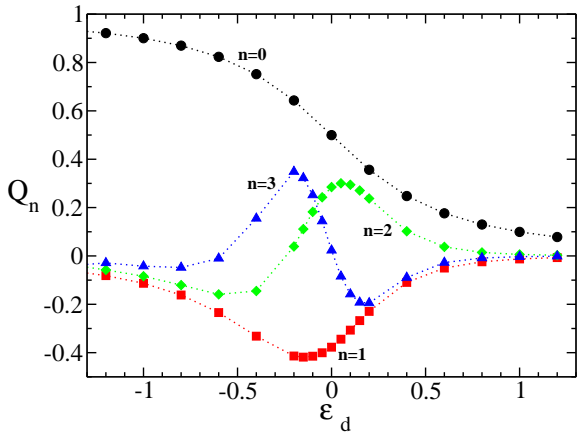}
    \caption{\label{fig:exactQn} $Q_n$ as a function of $\epsilon_d$ for $n=0$, $1$, $2$ and $3$. The calculations are performed using a direct evaluations of the integrals 
    in Eq.~\eqref{eq:basic} using the Simpson rule for $t=20$, $\gamma=0.5$ and $T=0$.}
\end{figure}

Knowing how to get the non-interacting Green's functions, we are now
ready to calculate the perturbation series. A first, rather naive, technique would consist in
calculating the integrals in Eq.~\eqref{eq:basic} using a simple discretization scheme (Simpson in our case).
Only the first few orders can be obtained that way, at large computational cost. 
Nevertheless, it is rather instructive and also serves as a check for the QMC algorithms discussed in the next section.
We focus on model A and compute the local charge $Q(U)=\langle n_\uparrow + n_\downarrow \rangle$ at various orders in $U^n$,
\be
\label{eq:Qserie}
Q(U)=\sum_{n=0}^{+\infty} Q_n U^n
\ee
Fig.~\ref{fig:exactQn} shows the resulting $Q_n(\epsilon_d)$ for $n=0...3$. With a parallel implementation,
results for $Q_4$ can also be obtained (not shown) at important computational cost and $Q_5$ is prohibitive.
All these results will be reproduced using the quantum Monte-Carlo sampling with a tiny fraction of the computational
time. Note that the stationary results obtained for $Q_n$ do not mean that the series Eq.~\eqref{eq:Qserie} is convergent,
but only that its coefficients are well defined.

\begin{figure}[h]
    \centering
    \includegraphics[width=8cm]{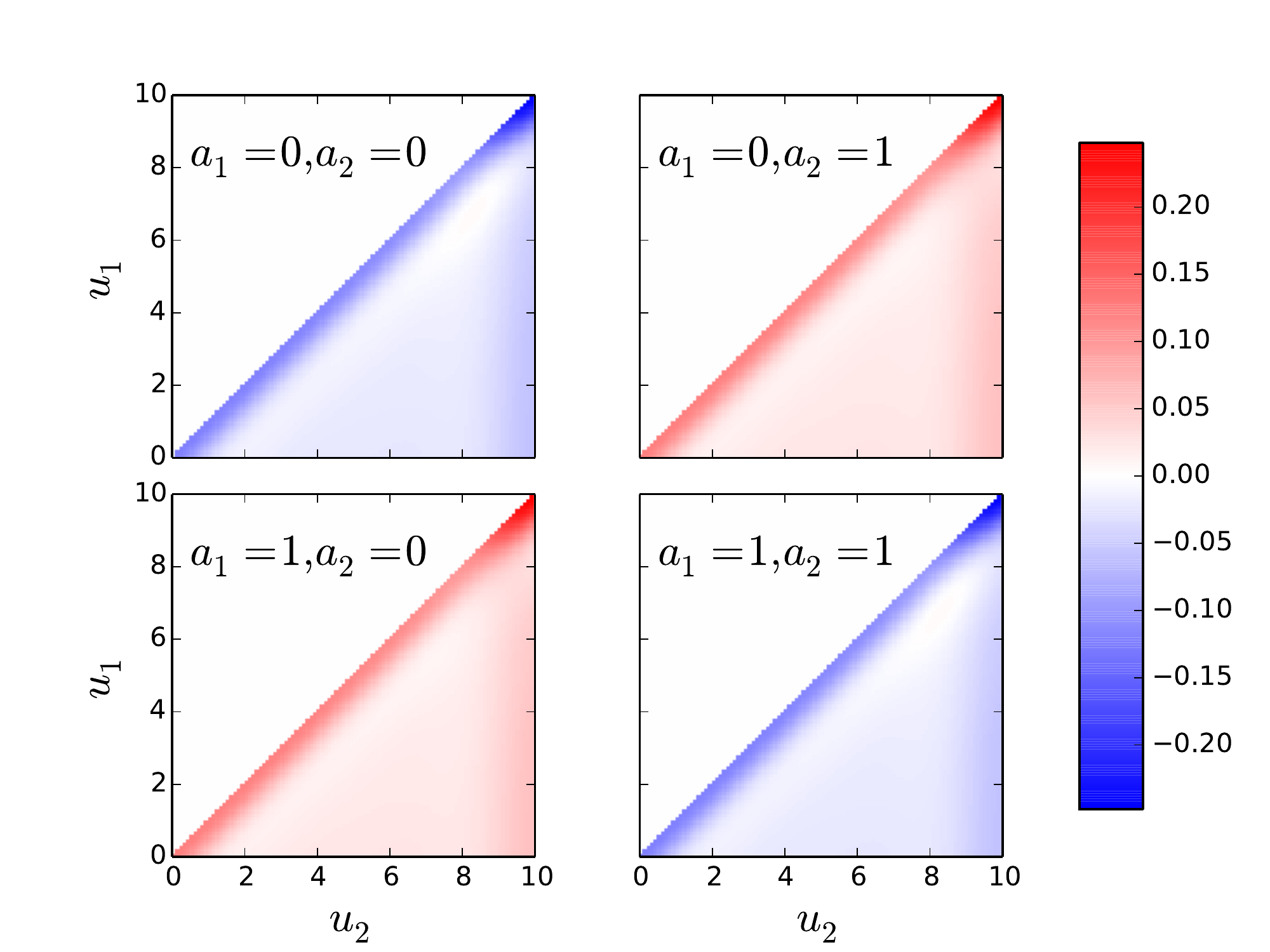}
    \caption{\label{fig:M} Colorplot of the integrand  of $Q_2$ as a function of the two times $u_1$ and $u_2$ for model A  with $\mu_L=\mu_R=0$, $\epsilon_d=0$, $T=0$ and $t=10$. 
    The four panels correspond to the 4 possible values of the two Keldysh indices $a_1$ and $a_2$. The explicit form of the integrand is $f(u_1,u_2,a_1,a_2)= -\Im m (-1)^{\sum_i a_i}  \det \nd{M}_{2}(u_1,u_2,a_1,a_2)$.
    }
\end{figure}

\begin{figure}[h]
    \centering
    \includegraphics[width=8cm]{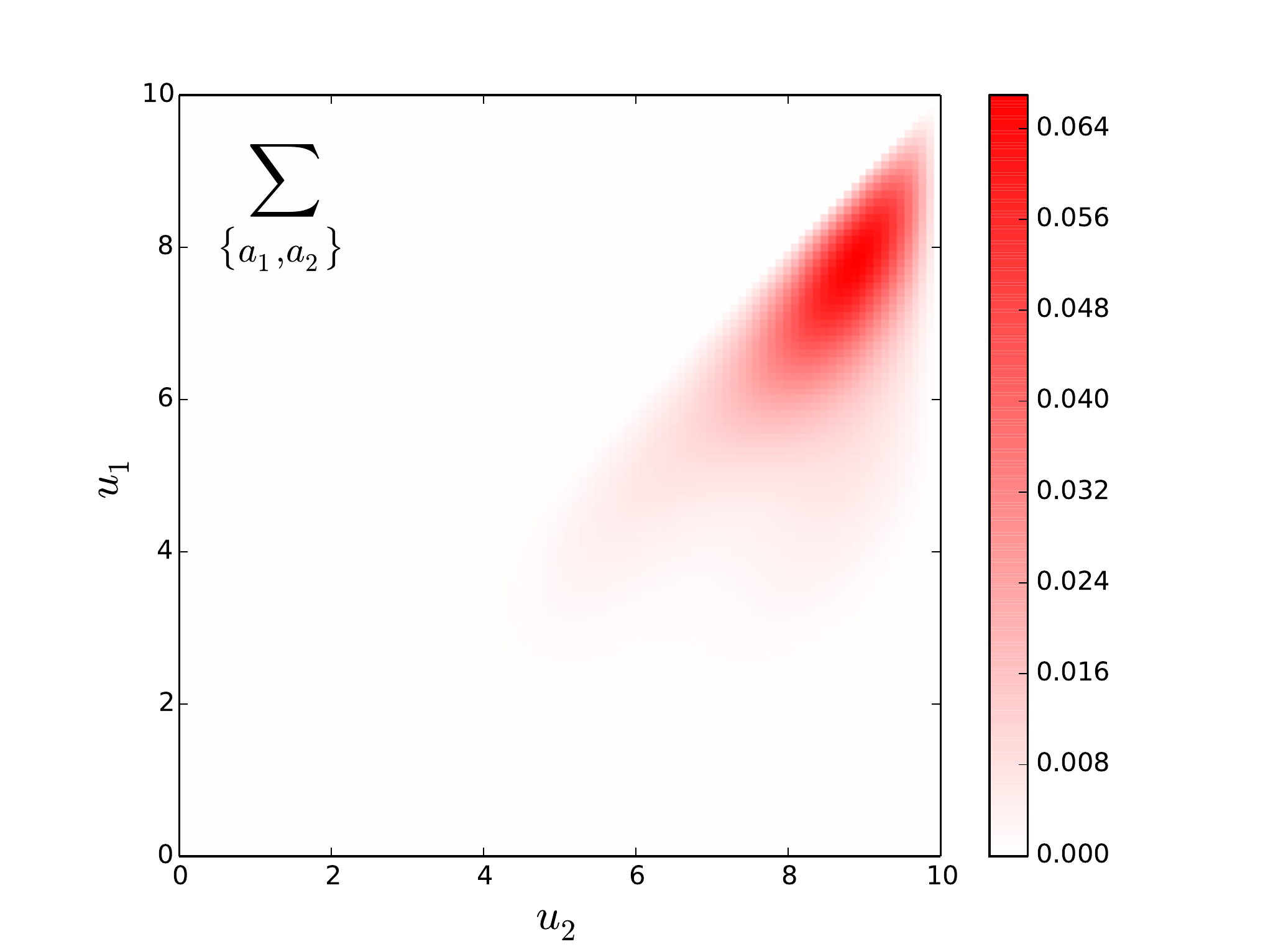}
    \caption{\label{fig:sumM} Same parameters as in Fig.~\ref{fig:M} but the integrand has now been summed over Keldysh indices. The colorplot represents $f(u_1,u_2)= i \sum_{a_1,a_2} (-1)^{\sum_i a_i}  \det \nd{M}_{2}(u_1,u_2,a_1,a_2)$
 ($f$ is real). Note that the integrand is now real, positive and concentrated around $u_1 = u_2 = t$. }
\end{figure}

\begin{figure}[h]
    \centering
    \includegraphics[width=8cm]{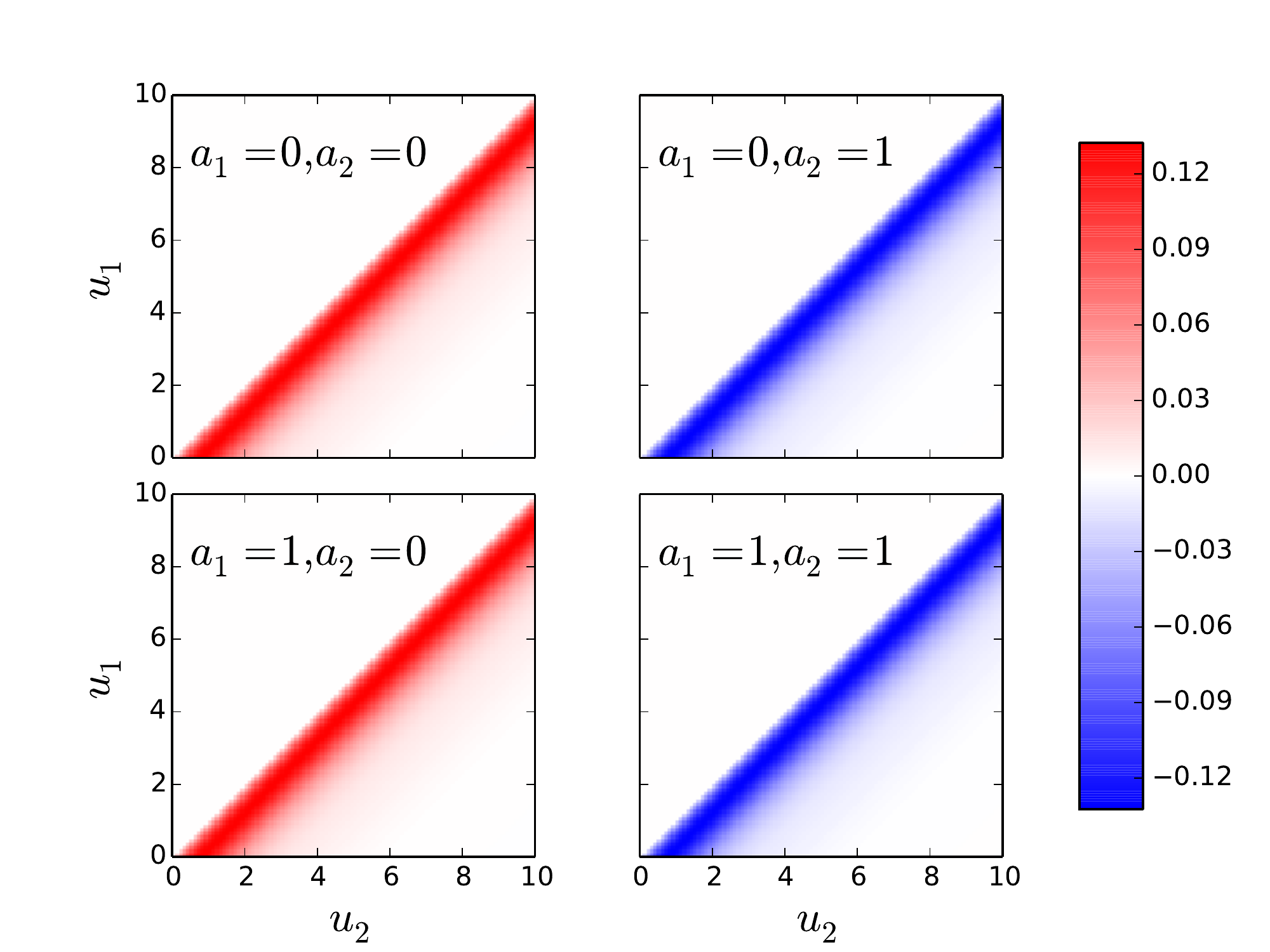}
    \caption{\label{fig:P} Same parameters as in Fig.~\ref{fig:M} but the integrand now uses the matrix $\nd{P}_2$ instead of $\nd{M}_2$, i.e. is associated to the partition function instead of an observable. The colorplot represents $f(u_1,u_2,a_1,a_2)= -\Im m (-1)^{\sum_i a_i}  \det \nd{P}_{2}(u_1,u_2,a_1,a_2)$
    }
\end{figure}

It is very instructive to have a look at the quantity which is actually integrated to obtain the $Q_n$. Fig.~\ref{fig:M} shows the integrand of $Q_2$ for the 4 values of the pair of
Keldysh indices $(a_1,a_2)$. We see that this integrand decays slowly as a function of $u_1-u_2$ and even more slowly as
$u_1$ or $u_2$ get away from the time $t$ where the charge is measured.  The sign of the integrand changes as one changes the
Keldysh indices. Fig.~\ref{fig:M} should be contrasted with Fig.~\ref{fig:sumM} which shows the same integrand but now summed over the four Keldysh indices. The integrand shown in Fig.~\ref{fig:sumM} now decays fast as $u_1$ or $u_2$ gets away from $t$. 
This observation can be proven and generalized for higher orders: the integrand decays to 0 when a group of $u_i$ is far from the 
time $t$ where the physical observable is measured, Cf. Appendix \ref{app:clustering}.
Finally, Fig.~\ref{fig:P} shows the same as Fig.~\ref{fig:M} but for the matrix $\nd{P}_2$ associated with the partition function. 
Note that for $\nd{P}_2$, the sum on the Keldysh indices simply vanishes, so there is no analogous Figure as  Fig.~\ref{fig:sumM} for $\nd{P}_2$.
In the next section, we will use these observations to design a better sampling strategy for the Monte-Carlo method.


\section{Quantum Monte-Carlo}
\label{sec:qmc}

The direct method of the previous section works in principle but is limited in practice to very small orders due
to its prohibitive computational cost. Stochastic methods, such as the Metropolis algorithm, can be extremely efficient at calculating integrals in high dimensions. In this section, we propose a new route to sample the interacting series by constructing a Markov process in the Fock configuration space (i.e. that not only samples the integrals themselves but also samples the various orders $n$ within one process).

\subsection{Sampling strategy}

Our algorithm is inspired by the conclusion of the previous section. 
It consists in {\it i)} sampling directly the physical quantity to be computed (and not the partition function, which is $Z=1$ anyway in the Keldysh formalism), 
and {\it ii)} summing explicitly over the Keldysh indices to restore unitarity (the symmetry between the two Keldysh contours) {\it for all configurations}.
Indeed, it is clear from the contrast observed between Fig.~\ref{fig:sumM} and Fig.~\ref{fig:P} that it is a much better choice, since the integration region 
is in the first case well localized around the time $t$ at which the quantity is computed.
Sampling  $\nd{P}_2$ would result in sampling large regions of the Fock space which are irrelevant to the actual observable.


We introduce:
\be
\label{eq:yeh}
{\bold\cal M}[{\cal C}_n]\equiv -i^{n+1} \sum_{\{a_i\}} (-1)^{\sum_i a_i} V({\cal C}_n)\det \nd{M}_{n}({\cal C}_n,\{a_i\})
\ee
which is {\it a real number}, as proven in Appendix \ref{real}. 
We construct a Markov process that samples the following density of probability,
\be
\label{eq:us}
 {\bold\cal P}[{\cal C}_n]=\frac{1}{Z_{\rm qmc}} \left|U_{\rm qmc}^n {\bold\cal M}[{\cal C}_n]\right|
\ee
where the denominator $Z_{\rm qmc}$ ensures that the probability is normalized. We use the notation $Z_{\rm qmc}$ and $U_{\rm qmc}$ to show explicitly that the "partition function" $Z_{\rm qmc}$ and interaction parameter $U_{\rm qmc}$ belong to the QMC technique. 
In particular, the physical value of the interaction $U$ can (and will) be distinct from the one(s) $U_{\rm qmc}$ used in the Monte-Carlo; 
also the physical partition function is $Z=1$ in the Keldysh formalism.

Introducing $O_n$, the contribution to order $U^n$ to the observable $O$ (see Eq.~\eqref{eq:O}):
\be
O(U) = \sum_n O_n U^n 
\ee
the terms of the perturbative expansion are given by
\be 
O_n/Z_{\rm qmc}= \ll \delta_{pn} \frac{1}{U_{\rm qmc}^p}   \frac{{\bold\cal M}[{\cal C}_p]}{\left|{\bold\cal M}[{\cal C}_p]\right|} \gg
\label{eq:obs}
\ee
where $\ll\dots\gg$ stands for the average over the probability ${\bold\cal P}[{\cal C}_n]$. All is left is
to construct a Markov process that samples this distribution. A key aspect of the approach is that $O_n$ is sampled for several (at least two) values of $n$ simultaneously: in the average $\ll\dots\gg$, $n$ varies between
$n_{\rm min}$ and $n_{\rm max}$. Introducing 
\be
c_n= \sum_{{\cal C}_n} \left|{\bold\cal M}[{\cal C}_n]\right|
\ee
we find that the probability $p_n$ to be in the order $n$ (in practice the fraction of the Monte-Carlo spent in configurations at order $n$) is 
\be
p_n=c_nU_{\rm qmc}^n /Z_{\rm qmc}
\ee
The normalization of the total probability $
 \sum_n p_n=1$ provides the partition function in terms of the $c_n$, $Z_{\rm qmc}=\sum_n c_n U_{\rm qmc}^n$. 
 Note that the $c_n$ are by definition independent of the QMC technique used to calculate them and in particular of the value of $U_{\rm qmc}$. 
 $c_0$ is simply given by the non-interacting value of the observable: $c_0=|g^<_{ij}(0)|$. 
The last item to introduce is the probability $q_n$ for the fluctuating sign in Eq.~\eqref{eq:obs} to be $+1$ (and $(1-q_n)$ to be $-1$). Note that $ {\bold\cal M}[{\cal C}_p]/\left|{\bold\cal M}[{\cal C}_p]\right|=\pm 1$ is always real (Cf Appendix \ref{real}) so that we only average fluctuating signs, not phases. We note $q_n=(1+s_n)/2$ so that $s_n$ is the average sign at a given order. $s_n$ and $p_n$ are the direct outputs of the computations. With these notations,
$\ll \delta_{pn}  {\bold\cal M}[{\cal C}_p]/\left|{\bold\cal M}[{\cal C}_p]\right|\gg= s_n p_n$ and one finally arrives at
\be 
O_n = c_n s_n
\ee
and
\be 
\frac{c_{n+1}}{c_n}= \frac{1}{U_{\rm qmc}}\frac{p_{n+1}}{p_n}
\label{eq:ratios}
\ee
which relates the observable ($O_n$) to the output of the computations ($p_n,s_n$). As always, Monte-Carlo
computations only provide {\it ratios } between quantities, see Eq.~\eqref{eq:ratios}. Here, we use our knowledge of the non-interacting observable (hence of $c_0$) to obtain the $c_n$ and finally the $O_n$
recursively. Eq.~\eqref{eq:ratios} needs to be applied for all $n$ up to the maximal order needed, but
its evaluation for various $n$ {\it needs not to} be done within the same QMC run.

\subsection{Moves and detailed balance}
Before we can actually perform calculations, we are left with a last task: designing a random walk that actually samples
Eq.~\eqref{eq:us} with $n$ varying between $n_{\rm min}$ and $n_{\rm max}$. 
This step is very similar to the construction of other continuous-time QMC. We introduce two sorts
of moves: the moves where we increase $n$ by one unit by adding one vertex and the moves where one vertex is deleted.
The algorithm to obtain the configuration ${\cal C}_n (i+1)$ at step $i+1$ from the configuration ${\cal C}_n (i)$ goes as follows:

{\bf (i) Move selection.} We choose the move $n\rightarrow n+1$ ($n\rightarrow n-1$) with probability $p_\uparrow$ ($p_\downarrow$) with $p_\uparrow + p_\downarrow =1$.
When $n=n_{\rm min}$ ($n_{\rm max}$) we have $p_\uparrow = 1$ ($p_\downarrow = 1$) otherwise we typically
choose $p_\uparrow = p_\downarrow = 1/2$.

{\bf (ii) Move $n\rightarrow n+1$.} 
We select $u_{n+1}$ uniformly in $ [0,t]$. We select $(i_{n+1},j_{n+1},k_{n+1},l_{n+1})$ uniformly among the $N_V$
different terms $V_{ijkl}$. The overall probability to propose the move is
\be
W_\uparrow du_{n+1}= \frac{p_\uparrow du_{n+1}}{t N_V}
\ee

{\bf (iii) Move $n\rightarrow n-1$.} We select the vertex to be removed uniformly between $[1\dots n]$. The overall probability to propose the move is
\be
W_\downarrow = \frac{p_\downarrow}{n}
\ee

{\bf (iv) Detailed balance.} We choose the acceptance probability $q_\uparrow(n)$ and $q_\downarrow (n)$ so that it satisfies the detailed balance equation
\begin{multline} 
W_\uparrow(n)du_{n+1} q_\uparrow(n) P[{\cal C}_n] \prod_{i=1}^n du_i=  \\
W_\downarrow(n+1) q_\downarrow (n+1) P[{\cal C}_{n+1}] \prod_{i=1}^{n+1} du_i
\end{multline} 
Using the Metropolis algorithm this leads to 
\begin{align}
&q_\uparrow(n)= {\rm min } \left(  \frac{ W_\downarrow(n+1) P[{\cal C}_{n+1}]}{W_\uparrow(n) P[{\cal C}_n]} , 1\right) \\
&q_\downarrow (n)= {\rm min }\left( \frac{W_\uparrow(n-1) P[{\cal C}_{n-1}] }{W_\downarrow(n) P[{\cal C}_{n}]}  1\right)
\end{align}
Note that, as usual in continuous-time quantum Monte-Carlo methods \cite{Prokofiev96, rubtsov2005,werner2006,gull2008}, 
the factor $du_{n+1}$ is present on both sides of the detailed balance equation and eventually drops, so that
the Monte-Carlo can  be performed directly in the (time) continuum.

\subsection{Remarks}
We now have a complete practical scheme for calculating many-body perturbation to a given observable.
Before we embark in concrete examples, let us make a few remarks.

\subsubsection{Comparison with the sampling of the partition function}
Our sampling strategy should be contrasted with the usual approach, used for instance in Ref.~\onlinecite{werner2010},
which has its origin in the imaginary-time techniques and where the (density of) probability ${\bold\cal P}[{\cal C}_n,\{a_i\}]$ to be in the configuration ${\cal C}_n$ 
with the Keldysh indices $\{a_i\}$ is given by
\be
\label{eq:rubtsov}
 {\bold\cal P}[{\cal C}_n,\{a_i\}]=\frac{1}{Z_{\rm qmc}} \left|U_{\rm qmc}^n V({\cal C}_n)\det \nd{P}_{n}({\cal C}_n,\{a_i\})\right|
\ee
i.e. one samples the matrix $\nd{P}_{n}$ (instead of $\nd{M}_{n}$) and one samples the
Keldysh indices (instead of summing on them exactly and explicitely). In this scheme,
an observable $O$ (say the charge $Q$, possibly resolved in spin) is given by
\be 
O(U_{\rm qmc})=Z_{\rm qmc} \ll (-1)^{\sum_i a_i}  \frac{\det \nd{M}_{n}}{|\det \nd{P}_{n}|} \gg
\ee
while the partition function $Z=1$ is given by
\be 
\label{eq:sgn}
1= Z_{\rm qmc}\ll (-1)^{\sum_i a_i}  \frac{\det \nd{P}_{n}}{|\det \nd{P}_{n}|} \gg
\ee
Constructing a Markov process that samples Eq.~\eqref{eq:rubtsov} can be done similarly to the construction presented in the previous subsection. The observable $O(U_{\rm qmc})$ can be estimated by taking the ratio of the above two equations.

Although this approach has shown some success, one can make the following remarks.

(1). The sign of $\det \nd{P}_{n}$ can fluctuate strongly so that the statistical average in Eq.~\eqref{eq:sgn} is very small and $1/Z_{\rm qmc}$ suffers from a very bad signal to noise ratio. This is the sign problem that plagues Quantum Monte-Carlo techniques for fermions. Eq.~\eqref{eq:zero} indicates that this problem is most probably worse in presence of the Keldysh indices where one expects wildly fluctuating signs.
This is shown by the data in Fig.~\ref{fig:P}.

(2). It is not guaranteed that the most probable configurations sampled by Eq.~\eqref{eq:sgn} are also the ones that contribute most to $O(U_{\rm qmc})$. On the contrary: the determinant of $\nd{P}_{n}$ depends only on the relative positions of the $u_i$ with respect to the others (it is essentially a sum of terms of the form $g^c_{ij}(u_1-u_3)g^c_{kl}(u_2-u_3)\dots g^c_{pq}(u_n-u_6)$) and not at all of $t$. The integrals contributing to $O(U_{\rm qmc})$ on the other hand decay when the $u_i$ get away from $t$. Hence for large times, the above scheme samples values of the $u_i$ very far from $t$ which therefore contribute very little to the actual observable. This was shown explicitly in  Fig.~\ref{fig:M}, Fig.~\ref{fig:sumM} and Fig.~\ref{fig:P}.

(3). The signal to noise ratio usually deteriorates rapidly with $t$ in these algorithms making it difficult to reach the stationary regime.

\subsubsection{Role of the explicit sum over the Keldysh indices}

The sampling of $Z=1$ discussed above does not preserve the symmetry between the two parts of the Keldysh contour for a given configuration (it preserves it in average, as it should): in the Keldysh formalism, one starts from a non-interacting density matrix, switches on the interaction for some time $t$, measures the observable, then unwinds the effect of the interaction until one is back to the original density matrix. Here however, a given configuration might have a few Keldysh indices on one branch and the rest on the other, meaning that a typical configuration does not enforce the symmetry of Keldysh indices, which reflects unitarity.
From a different perspective, the corresponding expansion includes all the Feynman diagrams,
including the {\it disconnected } diagrams although they have a vanishing contribution to the observables.

In our scheme in contrast, we explicitly sum over all Keldysh indices. Eq.~\eqref{eq:zero} indicates that all contributions from disconnected diagrams explicitly drop from the calculation. We calculate only connected diagrams which should be advantageous. From a technical perspective, one finds that the quantity ${\bold\cal M}[{\cal C}_n]$ is always real (not complex, see Appendix \ref{real}) so that one averages signs instead of phases. One expects that the resulting potential sign problem is milder than the so-called phase problem which originates from averaging a random phase. 

However, our scheme has an  obvious drawback: one evaluation of ${\bold\cal M}[{\cal C}_n]$ corresponds to the calculation of $2^n$ determinants, so that this algorithm complexity now increases exponentially with the maximum order $n$. We show below results for up to $n=15$ and it is reasonable to expect that one could calculate up to $n=20$. Usual algorithms to calculate determinants have complexities which scale as $n^3$. We show in appendix \ref{app:Gray} that the fast updates of the determinants (with complexities that scale as $n^2$) can easily be extended to the present case using Gray code, so that the overall complexity of our algorithm scales as $2^n n^2$. Actually, "mirror" Keldysh configurations have equal contributions (see Appendix \ref{real}) so that only $2^{n-1}$ configurations need to be included. Overall, we shall see  that the additional computational complexity is more than compensated by the important gain in signal to noise ratio.

\subsubsection{Statistical errors and the sign problem}

In practice, we calculate the moments $O_n$ sequentially with a separate QMC computation for each value of $n$
(typically with $n_{\rm min}=n-1$, $n_{\rm max}=n$ or  $n_{\rm min}=0$, $n_{\rm max}=n$) so that one gets
a fully controlled (optimally flat) histogram of the various orders.
Starting from $c_0$ (known without error), we iteratively compute $c_n$ from different runs that use different interaction strengths $U_n$,

\be 
\frac{c_{n}}{c_{n-1}}=A_n \equiv \frac{p_{n}(U_n)}{U_n p_{n-1}(U_n)}
\ee
One can use for instance $U_n$ such that $p_n=p_{n-1}=1/2$ (this is always possible but not strictly necessary as long as $p_n/p_{n-1}$ remains of order unity).

Note that a naive scheme where one would try to evaluate all the values of $c_n$ in a single run would run
into an artificial difficulty: in a single run, the histogram is usually sharply peaked around one value of $n$ and one cannot get a good statistics both for $n=1$ and $n=n_{\rm max}$. A small statistics in, say $n=1$ leads to a large error in the estimate of $p_1$ which further corrupts the evaluation of all $c_n$, hence $O_n$. 

Coming back to our scheme, the error made on the estimate of $A_n$ is bounded by $\delta A_n/A_n \le 2/\sqrt{N_\#}$ where $N_\#$ is the number of independent points. Hence we find that the (one standard deviation) error on $c_n$ is bounded by
\be
\frac{\delta c_n}{c_n} \le \frac{2n}{\sqrt{N_\#}}
\ee
which can be controlled to arbitrary precision provided $n \ll \sqrt{N_\#}$. Hence, the calculation of the $c_n$, which involves only positive numbers, can be done with extremely good accuracy. 

The limitation to the precision -- the sign problem -- takes its origin in the average $s_n$ of the sign contained in Eq.~\eqref{eq:obs}. Note that in contrast to other techniques, this sign is here in the {\it numerator} and is not present in the denominator, i.e. a small sign does not necessarily mean a sign problem: it can simply mean a small value of $O_n$. This analysis is close to e.g. Refs \onlinecite{Prokofiev08polaron, DiagQMC, KozikEPL10, van2012NatPhys}. 
The error made on the sign $s_n$ is given by a Bernoulli law ($+1$ with probability $q_n$, $-1$ with probability $1-q_n$), hence is given by $2\sqrt{[q_n(1-q_n)]/N_\#}$ which is always smaller than \be
\delta s_n \leq \frac{1}{\sqrt{N_\# }}
\ee(the upper bound is reached when the sign $s_n$ becomes small). Putting everything together, $O_n= c_n s_n$ implies
that $|\delta O_n/O_n| = \delta s_n / |s_n| + \delta c_n / c_n$ and we arrive at the relative error,
\be 
\left|\frac{\delta O_n}{O_n}\right| \le  \frac{1}{|s_n|\sqrt{N_\# }} + \frac{2n}{\sqrt{N_\#}}
\ee
In this form, it seems that the smaller the sign, the larger the error, hence the sign problem. However,
one must remember that while $c_n$ and $s_n$ depend on the QMC algorithm, their product $c_n s_n = O_n$
does not, so that the error can be recast into
\be 
\left|\frac{\delta O_n}{O_n}\right| \le  \frac{c_n}{|O_n|\sqrt{N_\# }} + \frac{2n}{\sqrt{N_\#}}
\ee
In this second form, it becomes apparent that a bad sign problem (small $s_n$) is 
equivalent to a bad sampling choice which leads to a large $c_n$. The behaviour of the error is therefore intimately linked with the growth of $c_n$ with $n$ which itself depends strongly on the actual probability sampled. For instance, if one samples the sum over Keldysh indices (instead of summing explicitely over the indices as we do), one gets identical $O_n$ but much larger $c_n$ and consequently much smaller $s_n$.
In fact the $c_n$ would increase by more than a factor $2^n$ so that the overall method would be much less efficient than the one proposed here.
The global relative
error therefore contains three contributions,
which reflect respectively the total computing time ($1/\sqrt{N_\# }$), the intrinsic physics of the problem ($O_n$) and the quality of the choice of the sampling procedure ($c_n$). From the above analysis, the colorplots shown in Fig.~\ref{fig:M}, Fig.~\ref{fig:sumM} and Fig.~\ref{fig:P}
take a different meaning. Indeed, one can see that $c_2$ (the integral of the function displayed in Fig.~\ref{fig:sumM})
is rather small: the function decays rather quickly when $u_1,u_2$ get away from $t$. If on the other hand we have chosen to sample the Keldysh indices in addition (as in the standard schemes), then $c_2$ would have been the sum of the integral of (the absolute value of) the different panels of Fig.~\ref{fig:M}. One immediatly realizes that the signal to noise ratio would have been much smaller.

\subsubsection{Convergence of the interacting series}

Our approach separates the calculation of the $O_n$ from the study of (the convergence of) the series
$O(U) = \sum_n O_n U^n$ itself.  This could be used to obtain the full $U$ dependence of the observable, but more importantly, it allows one to disentangle physical aspects (for instance the convergence or lack of of the series) from technical ones (the calculation of its elements). The convergence of the series will be discussed next. In principle, one could use various resummation procedures such as Pade approximant, Lindel\"of analytical continuation and/or Borel resummation in order to extrapolate the series from its first coefficients. An example of such a procedure is given in the appendices.

It is important to note already at this stage that the parameter $\alpha$, as introduced in model B, plays
a crucial role in the algorithms sampling the partition function at equilibrium \cite{rubtsov2005} as only special values of
$\alpha$ are free of the sign problem. It is also known that the typical perturbation order in those algorithms is strongly reduced by using 
the best value of $\alpha$. 
We shall find in the next section that the parameter $\alpha$ has a drastic influence on the convergence of the series $\sum_n O_n U^n$ but not on the actual calculation of the $O_n$ itself.

\section{First results: analysis of the series convergence}
\label{sec:bare}
\begin{figure}[h]
    \centering
    \includegraphics[width=8cm]{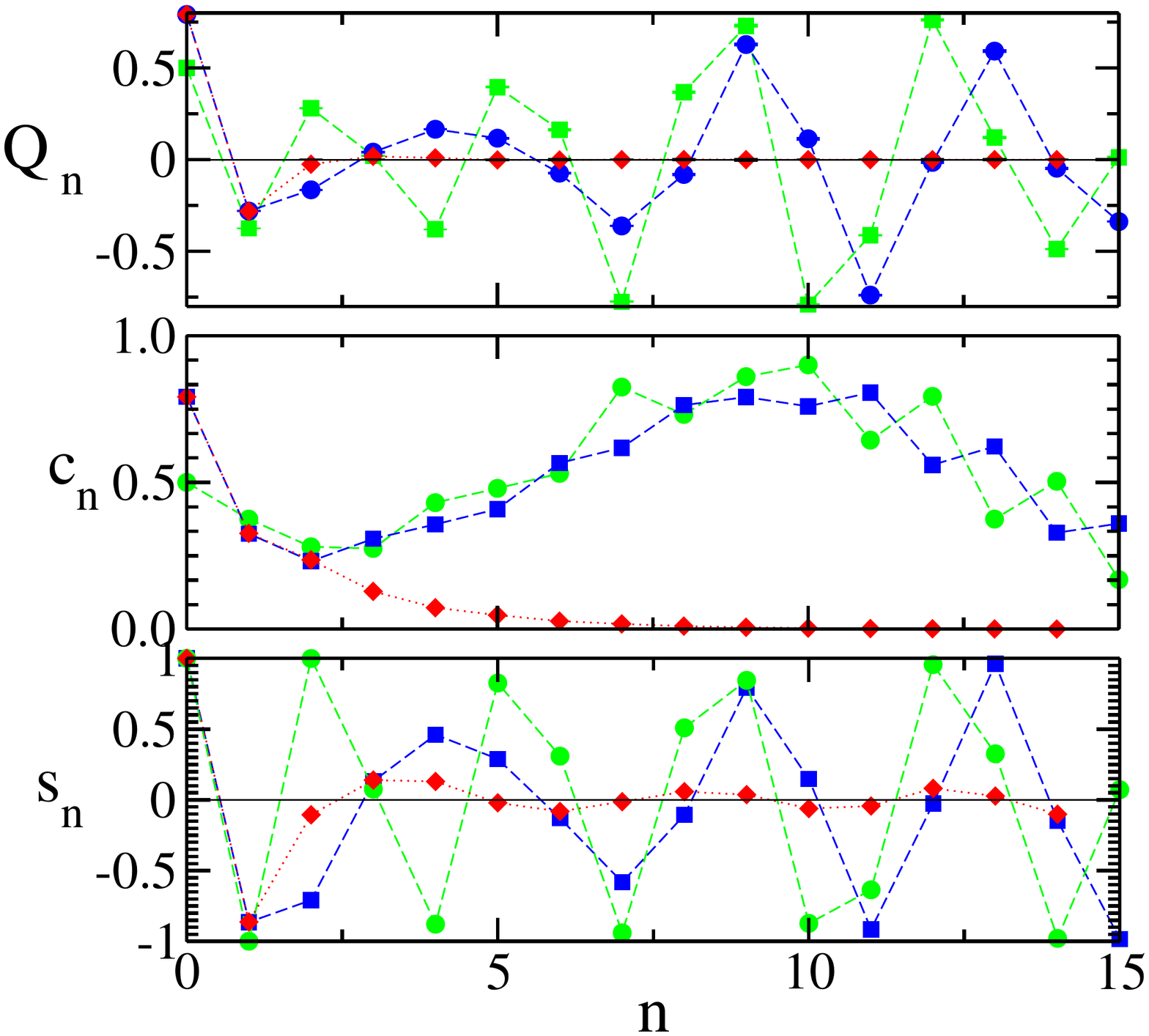}
    \caption{\label{fig:modelbare} 
    QMC results for model A at $\gamma=1/2$, $T=0$ and $t=10$ for $\epsilon_d=0$ (green squares) and $\epsilon_d=-0.5$ (blue circles), as a function of the order $n$. Red diamonds: model B with $\gamma=1/2$, $T=0$ and $t=10$ for $\epsilon_d=-0.5$ and $\alpha=0.5$. 
    Top panel: $Q_n$, central panel: $c_n$ and bottom panel: average sign $s_n$.
    }
\end{figure}

\begin{figure}[h]
    \centering
    \includegraphics[width=8cm]{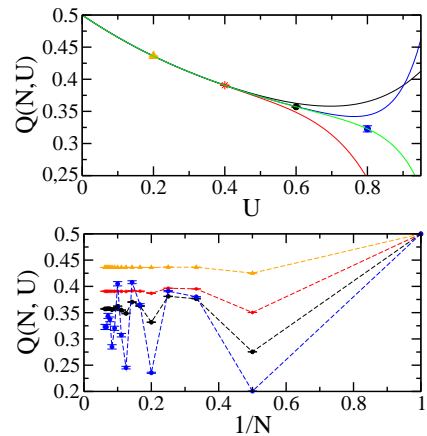}
    \caption{\label{fig:modelbare2} QMC results for model A at $\gamma=1/2$, $T=0$ and $t=10$.
    Top panel: charge $Q(N, U)$ as a function of $U$, for different $N$ and for $\epsilon_d=0$: $N=5$ (Black), $N=7$ (red),
     $N=13$ (blue) and $N=15$ (green). Bottom panel: $Q(N, U)$ as a function of $1/N$ for different $U$: $U=0.2$ (Orange),
     $U=0.4$ (red), $U=0.6$ (black) and $U=0.8$ (blue).
    }
\end{figure}

\subsection{Bare results}

As a first application, we compute the interaction corrections to the charge $Q$
on the impurity in model A. Fig.~\ref{fig:modelbare} shows the 
$Q_n$ for $n$ up to $n=15$ as well as the corresponding $c_n$
and $s_n$. We find that the magnitudes of the $Q_n$ do not appear to decrease with $n$ but rather
remain of order unity $|Q_n| \approx 1$. This implies (Hadamard's theorem) that the series has an apparent convergence radius of order unity (apparent because it relies on extrapolating the behaviour of the first known coefficients to large orders). We can already notice that the curve for model B ($\alpha=1/2$) has a very different behaviour with a fastly decreasing $c_n$ hence $Q_n$;
this aspect will be discussed later in the text.
Fig. \ref{fig:modelbare2} shows the truncated series
\be
\label{eq:Qtronc_serie}
Q(N,U)=\sum_{n=0}^{N-1} Q_n U^n
\ee
as a function of $U$ (upper panel) and $1/N$ (lower panel). We find a nice convergence for $U<U^*\approx 0.6$
but a divergence beyond, i.e. we cannot access the physics beyond $U=0.6$ ($U\approx\Gamma/2$) by simply summing the series. 
This divergence has nothing to do with the QMC technique which is just a way to calculate
the $Q_n$ - it belongs to the physics of the problem. In the following, we will discuss two ways
to bypass this problem: performing the interaction expansion from a different starting point 
and resumming the series by moving its singularities away from the expansion point.

\subsection{Hartree-Fock series}
\begin{figure}[h]
    \centering
    \includegraphics[width=8cm]{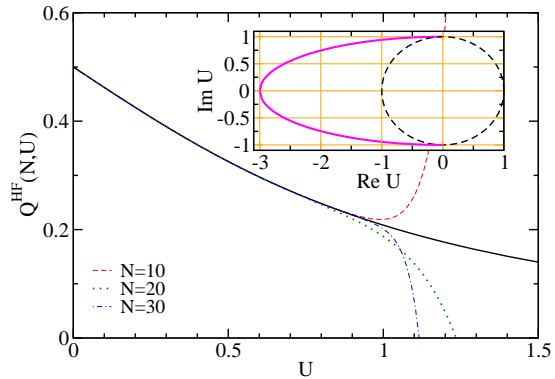}
    \caption{\label{fig:HF} Hartree-Fock series of model A as a function of $U$: Exact curve $Q^{HF}(U)$ (thick line) and partial sums $Q^{HF}(N,U)$ for $N=10,20,30$, $\epsilon_d=0$ and 
 $\gamma=1/2$. The series has a divergence at $U=1$. The inset shows the analytical structure of $Q^{HF}(U)$ in the complex plane 
 $({\rm Re}\ U, {\rm Im}\ U)$: branch cut of $Q^{HF}(U)$ (thick magenta line) and convergence radius of the Hartree-Fock series (dashed line).
    }
\end{figure}

Before going on with the QMC results, let us analyse a simpler,  approximate, series
which is obtained from the Hartree-Fock approximation (which reduces to Hartree for model A). 
This series will be useful in identifying a possible source for the observed apparent radius of convergence.
Introducing,
\be
\label{eq:hf}
g^R(E)=\frac{1}{E -\epsilon_d -\gamma^2 (E +i\sqrt{4-E^2})}
\ee
the Fourier transform of the retarded non-interacting Green's function $g^R_{00}(t)$, the 
non-interacting charge at equilibrium and $T=0$ takes the form
\be
Q_0=\frac{1}{\pi} \int_{-2}^0  dE \ \ {\rm Im}\  g^R(E)
\ee
In the Hartree approximation, one replaces the on-site energy by its mean-field value.
The fully self-consistent Hartree would be defined as $Q^{HF}(\epsilon_d,U)=Q_0[\epsilon_d + U Q^{HF}(\epsilon_d,U) ]$.
Here however, we restrict ourselves to summing the ladder of tadpole diagrams which is sufficient to illustrate our point:
$Q^{HF}(\epsilon_d,U)=Q_0[\epsilon_d + U Q_0(\epsilon_d) ]$. The corresponding series is given by $Q^{HF}(U)=\sum_n Q^{HF}_n U^n$
with
\be
Q^{HF}_n= \frac{Q_0^n}{\pi} \int_{-2}^0  dE \ \ {\rm Im}\  [g^R(E)]^n
\ee
Note that the first two moments are the exact ones: $Q^{HF}_0=Q_0$ and $Q^{HF}_1=Q_1$.
On the other hand, from the above construction Eq.(\ref{eq:hf}), we find that  $Q^{HF}(U)$ has branch cuts in the complex plane, given by
\be
E -\epsilon_d -U Q_0(\epsilon_d) -\gamma^2 (E \pm i\sqrt{4-E^2})=0 \ \ {\rm for } \ \ -2\le E\le 0
\ee
so that one expects the series in powers of $U$ to have a finite radius of convergence given by the branch closest to $U=0$, i.e. equal to unity (see the inset of Fig.~\ref{fig:HF}). Indeed, Fig.~\ref{fig:HF} shows an example of the partial sums which diverges around $U=1$. This is very reminiscent to what we have found for model A. This calculaiton illustrates in a simple and tractable approximation that a finite radius of convergence is due to the existence of singularities or branch cuts in the complex plane.

\subsection{Using a different non-interacting problem}
In this section, we show that by playing with the $\alpha$ parameter, one can greatly enhance the radius of convergance of the series hence access correlated regimes. Instead of starting the perturbation from the $U=0$ Hamiltonian, one can incorporate the mean-field treatment into the non-interacting Hamiltonian so that only the {\it fluctuations} of the interaction need to be taken into account in the perturbation. In fact, one can even push this idea further and 
add an arbitrary quadratic Hamiltonian to $\hnd{H}_0(t)$ and remove it accordingly from $\hnd{H}_{\rm int}(t)$. 
The idea is to start with one-body propagators as close as possible to the interacting ones, so that the role of the perturbation becomes very weak. 
Within our current implementation we can easily add an on-site potential $\delta \epsilon_d$ to the non-interacting Hamiltonian and use $\alpha=\delta\epsilon_d/\bar U$ in the perturbation where the new parameter $\bar U$ is our targetted value of the interaction (see the definition of model B in Eq.~\eqref{eq:modelB}). For  $U=\bar U$, one recovers the original model A for $t>0$, hence the corresponding results in the long-time limit. 

The $\alpha$ parameters are exactly the same as the ones used in the interaction expansion continuous-time quantum Monte-Carlo
introduced by Rubtsov \cite{rubtsov2005}, in equilibrium. In that algorithm, it was shown \cite{rubtsov2005} that the sign problem strongly depends on the value of $\alpha$. It also reduces the average order of perturbation of these QMC methods
Here, we will now show that {\it the apparent radius of convergence of the interaction expansion strongly depends on these $\alpha$ parameters}, 
and we will use this to our advantage.

\begin{figure}[h]
    \centering
    \includegraphics[width=7cm]{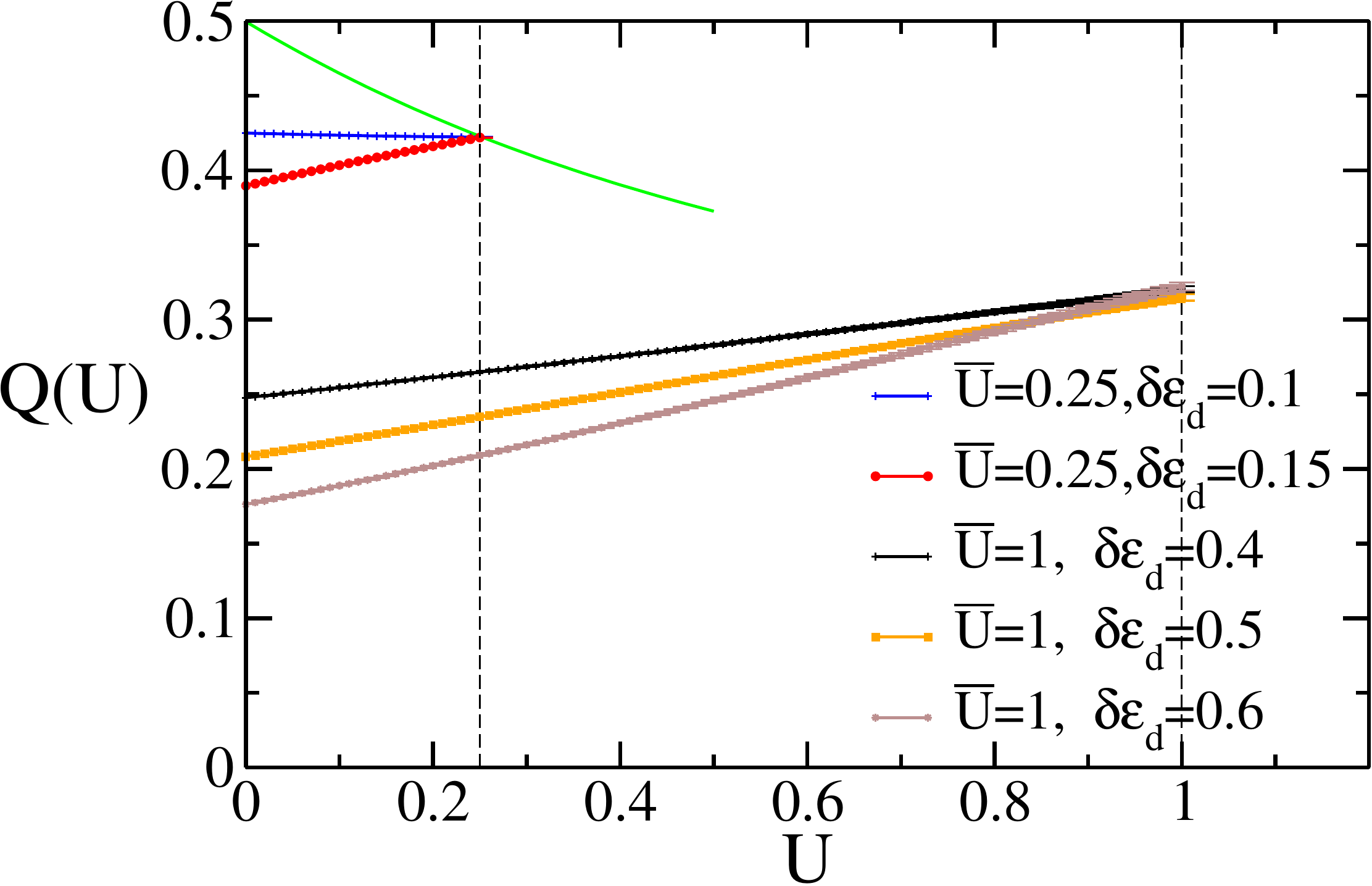}
    \caption{\label{fig:modelA_3} Role of the non-interacting Hamiltonian around which is done the expansion; $Q(U)$ at $\epsilon_d=0$ with an extra potential 
    $\delta\epsilon_d$ in the non-interacting Hamiltonian and a corresponding compensating term $\alpha=\delta\epsilon_d/\bar U$
    in the perturbation. 
    The green line corresponds to the original series with $\delta\epsilon_d=0$ in its regime of convergence. 
    All curves for $\gamma=1/2$, $T=0$ and $t=10$.}
\end{figure}

The technique is illustrated in Fig.~\ref{fig:modelA_3} first for a value of $U$ that could be reached with the initial approach ($U=0.25$)
and secondly for a value that could not be reached ($U=1$). We find that this approach works remarkably well: not only  we can
recover the former results, but we can also go to regimes that were not accessible before. As a self-consistency check, we find that the results do not depend on $\delta\epsilon_d$ for $U=\bar U$. Of course a disadvantage of this approach is that one must perform a separate computation for each value of $U$ needed.

\begin{figure}[h]
    \centering
    \includegraphics[width=7cm]{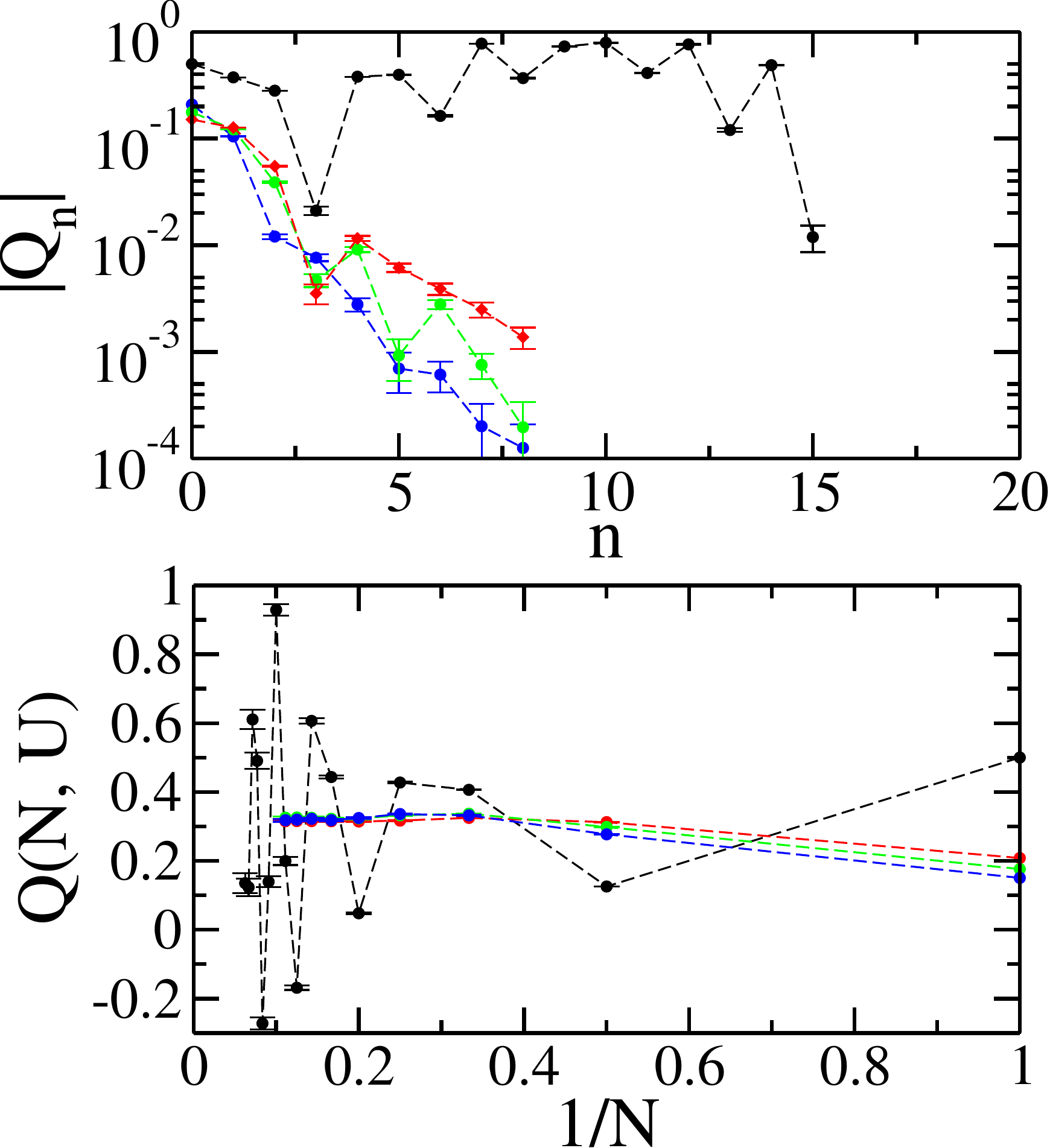}
    \caption{\label{fig:modelA_1} Top panel: absolute value of the charges $Q_n$ for different values of $\alpha$,$\delta\epsilon_d$, at $\epsilon_d=0$, $\gamma=1/2$, $T=0$ and $t=10$:
  $\alpha=\delta\epsilon_d=0$ (Black), $\alpha=\delta\epsilon_d=0.5$ (Blue), $\alpha=\delta\epsilon_d=0.6$ (Green),
    $\alpha=\delta\epsilon_d=0.7$ (Red).
    Bottom panel: corresponding dependence of the charge $Q(N,U=1)$ as a function of $1/N$.
    }
\end{figure}

\begin{figure}[h]
    \centering
    \includegraphics[width=7cm]{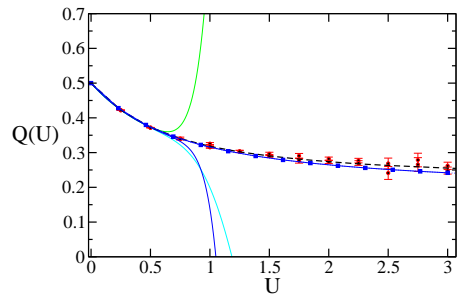}
    \caption{\label{fig:modelA_2} Charge $Q(U)$ for model A with $\epsilon_d=0$, $\gamma=1/2$, $T=0$ and $t=10$. 
    Red symbols correspond the use of $\alpha$ parameters (different symbols for the same value of $U$ correspond to different $\delta\epsilon$ or $\alpha$). Thin lines correspond to partial sum with the bare technique ($\alpha =0$): $Q(N=6,U)$ (cyan), $Q(N=7,U)$ (green) and $Q(N=8,U)$ (blue). Dashed line: Benchmark calculation performed with hybridization QMC in imaginary time. Blue squares: extrapolation using the homographic technique, see section\ref{sec:homo}.}
\end{figure}

The faster convergence of the new series can be seen in Fig.~\ref{fig:modelA_1} where we compute the corresponding
series $Q_n$ versus $n$ (upper panel) as well as the convergence of the partial sum $Q(N,U=\bar U)$ versus $1/N$
(lower panel). We find that only 2 or 3 orders are sufficient to obtain the exact result provided one uses a "starting point"
close enough to the final solution. A pragmatic way to perform the calculation is therefore to optimize the value of $\delta\epsilon_d$ so that the corresponding $Q_n$ decreases as fast as possible.
We collect the final curve $Q(U)$ for model A at $\epsilon_d=0$ (note that other values are equally accessible) in Fig.~\ref{fig:modelA_2} together with the original expansions. We find that $Q(U)$ decreases monotonously from $Q(0)=1/2$ to
$Q(U\gg 1)\approx 1/4$. For $U>2$ the model is already close to its large $U$ limit and fluctuations of charge are small.
Fig.~\ref{fig:modelA_2} is an important result of this paper and establishes that the regime of strong interaction can be reached
from a rather naive perturbative expansion. We note that the regime shown in Fig.~\ref{fig:modelA_2} is a quite difficult one for the technique. Indeed raising either the temperature and/or a bias voltage will result in Green's functions that
decay rapidly with time (exponentially as opposed to the algebric decay found at zero temperature) and therefore in smaller $c_n$
and better signal to noise ratio in the calculations. Fig.~\ref{fig:modelA_2} also includes a separate calculation performed with an
hybridization QMC technique in imaginary time (dashed line),
obtained with the algorithm introduced in Ref.~\onlinecite{werner2006}, implemented with the TRIQS package \cite{TRIQS}. We find a perfect match between the two techniques which serve as a validation of our technique and its implementation.

\subsection{Resummation technique: moving the singularities away}
\label{sec:homo}
Let us go back to our initial series for model A (with $\alpha=0$). Our
Hartree-Fock analysis suggests that the source of divergence lies in the complex plane (magenta line in 
Fig.~\ref{fig:HF}). However, the Anderson model does not display a transition (hence no singularity) on the real axis for $U$,
which suggests that one should be able to analytically continue our result for $U>1$.

There is a large body of literature dedicated to the study of diverging or even asymptotic series, see e.g. Refs \onlinecite{hardy,van2012NatPhys,mera2014} and various techniques can be used to extract useful physical information from them.
In Appendix \ref{app:lindelof}, we use the Lindel\"of method to extrapolate the series beyond the radius of convergence, for the charge $Q(U)$, and show that we can easily get the correct result for e.g. $U=1.5$.
Alternatively, one could use Borel resummation followed by a Pade fit (not shown). The Borel series, like
the Lindel\"of method, allows one to go beyond the initial radius of convergence, but not much beyond. It is also
rather impractical to do in a controlled way.

\begin{figure}[h]
    \centering
    \includegraphics[width=9cm]{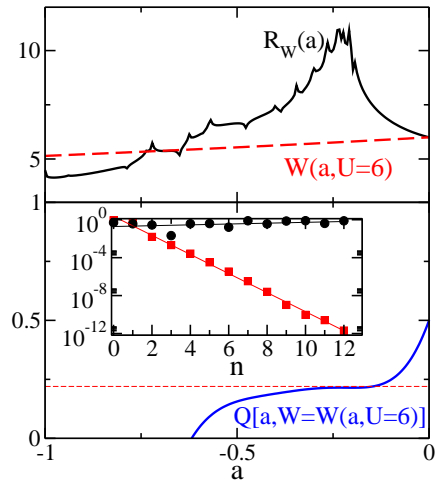}
    \caption{\label{homo} Homographic transformation of the model A series for $Q(U)$ (same data as Fig.\ref{fig:modelbare2}) with $b=6$. Upper panel: Effective radius $R_W(a)$ (ful black line) and homographic transformation $W(a,U=6)$ (dashed red line) as a function of position of the singularity $a$. Bottom panel: resummation result $Q(U=6)$ as a function of $a$. The dashed line indicates the exact result $Q\approx 0.22$. Inset: $\bar Q_n$ as a function of $n$ for $a=-0.25$ (squares) and the initial series (circles).}
\end{figure}

Here, we use an alternative route known as the Euler transform\cite{DombGreen}. The idea is to perform a meromorphic transformation
$W(U)$ that sends the singularities away from the expansion point while bringing the region of interest (real and positive $U$s) closer to zero (with $W(0)=0$). As an illustration, we choose the homographic transform
\be 
W(U) = \frac{b U}{U - a}
\ee 
The method is performed in two steps. First, one obtains the expansion for the inverse $U(W)$ of $W(U)$
(defined as $U[W(U)]=U$): $U(W)= U_1 W + U_2 W^2 + U_3 W^3 \dots$ and one constructs the (truncated) series for $Q(W) \equiv Q[U(W)]$. The series $Q(W)=\sum_n \bar Q_n W^n$ has a radius of convergence $R_W$ which may be much larger than the initial series if $a$ is close enough to the singularities of $Q(U)$. In a second step, one evaluates $Q(W)$ for $W=W(U)$.
If $R_W > W(U)$ the result will be convergent so that the figure of merit for this transformation is the ratio
$R_W/W(U)$. In practice, $R_W$ is obtained by performing a simple exponential fit for the $\bar Q_n  \propto R_W^{-n}$
(see the inset of Fig.\ref{homo} for an example). 
It is also very appealing conceptually and controlled by the figure of merit $R_W/W(U)$ (in practice $R_W/W(U)\approx 1.5$ gives very precise results with only 7-8 orders).

Fig.\ref{homo}  gives an example of the resummation for $U=6$ (much beyond the radius of convergence of the initial series). We find that the figure of merit reaches a rather high value $R_W/W(U)\approx 2$ which allows
one to obtain the exact result with only a few orders. The increase of the radius of convergence is actually quite dramatic. In the lower panel we plot the actual result obtained as a function of $a$ (the results do not depend on $b$). We find that, in the region where the figure of merit is high enough, one observes a nice plateau at the correct value. We have also reported
the full extrapolated $Q(U)$ curve in Fig.~\ref{fig:modelA_2} (blue squares) which is in very good agreement with our reference calculation. This method has two clear advantages over other resummation methods. First, the technique is controlled with the figure of merit. Second it can, in principle, works even in the strong coupling limit. For the case studied here, the point $U=\infty$ is mapped onto $W=b = 6$ which is well within 
our convergence radius $R_W(a=-0.25) \approx 12$ so that we can compute $Q(U=\infty) \approx 0.18$.

\section{More results: current and magnetization}
\label{sec:kondo}
In the previous section, we have computed  the charge of the impurity, a quantity that misses the important physics associated to spin fluctuations, the Kondo effect. The Kondo effect in quantum dots has been extensively studied and we refer to Ref.~\onlinecite{pustilnik2004,mora2014} for an account of the literature. Here, we merely aim at illustrating our method with calculations of current versus voltage and magnetization versus field characteristics.

\subsection{Current}

\begin{figure}[h]
   \centering
    \includegraphics[width=10cm]{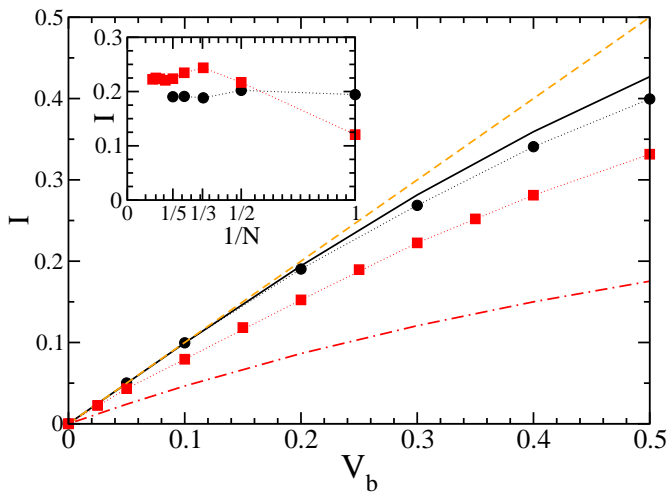}
   \caption{ \label{fig:IV} Current-voltage characteristics $I(V_b)$ for model B with $\alpha=0.5$, $\gamma=0.5$ and $T=0$.
   The current is in units of energy time $e/h$. Black line and black circles: particle-hole symmetric point 
   $\epsilon_d=0$ for $U=0$ and $U=1.2$ respectively. 
   Red dot-dashed line and red squares $\epsilon_d=-0.5$ for $U=0$ and $U=1.2$. 
   Dashed line: perfect transmission $I=V_b$. Inset: convergence of the results as a function of $1/N$ where $N$ is the total number of moments included for $V_b=0.3,\epsilon_d=-0.5,U=1.2$ (red squares) and
   $V_b=0.2,\epsilon_d=0,U=1.2$ (black circles)
   }
\end{figure}

In  Fig.~\ref{fig:IV}, we present some results  obtained for the current-voltage characteristics
of an Anderson impurity connected to two electrodes. 
 
Fig.~\ref{fig:IV} shows the resulting $I(V_b)$ characteristics for two values of 
$\epsilon_d$: $\epsilon_d=0$ which corresponds to the particle-hole symmetric case where the non-interacting
problem is at resonance; and $\epsilon_d=-0.5$ which breaks this symmetry and is non-resonant. We also
indicate the perfectly transmitting limit $I=V_b$ with the orange dashed line. The $\epsilon_d=0$ case is
already very interesting: at small bias, the transmission probability is unity because we are exactly at the resonance frequency of the impurity. When one increases the interaction strength, we also expect perfect transmission but for a totally different reason: the original resonance is now shifted to negative energies
($-0.6$ in this instance) but a new one, the Kondo resonance, starts to develop at the Fermi level ($V_b\le T_K$). In practice, we find that the $I_n$ are extremely small at small voltages so that the current is 
unaffected by the interaction. It is only at higher bias than interaction becomes relevant. It is interesting 
to note that this "Kondo ridge" which is a notoriously difficult regime appears here to be one of the most 
tractable ones.
The off-resonant case $\epsilon_d=-0.5$ corresponds to a situation where the non-interacting current is much 
smaller than the interacting one so that the perturbation series must build the Kondo resonance. We find that 
it does indeed build it as one recovers perfect transmission at low bias. Up to the accuracy of the calculations (the error bars are of the order of the symbol sizes), the result shown in Fig.~\ref{fig:IV}
is an exact solution of the non-equilibrium Anderson model, in its stationary regime, with and without particle-hole symmetry.

\begin{figure}[h]
   \centering
    \includegraphics[width=9cm]{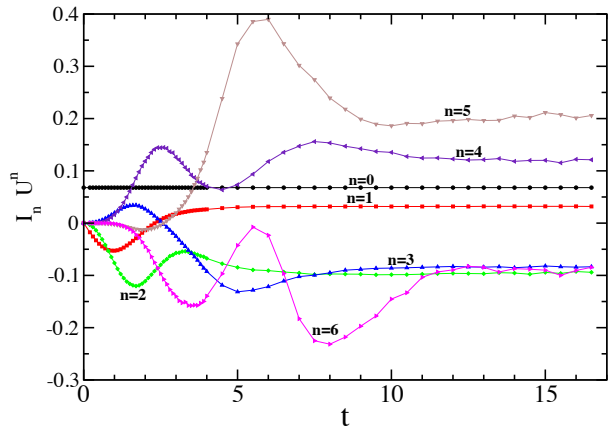}
   \caption{\label{fig:In} First 6 moments of the currents $I_n U^n$ versus time $t$. The moments have been rescaled with $U=3$ for visibility. Model B with $\alpha=0.5$, $\epsilon_d=0$, $\gamma=0.5$, $T=0$ and $V_b=0.5$.
   }
\end{figure}

\begin{figure}[h]
   \centering
    \includegraphics[width=9cm]{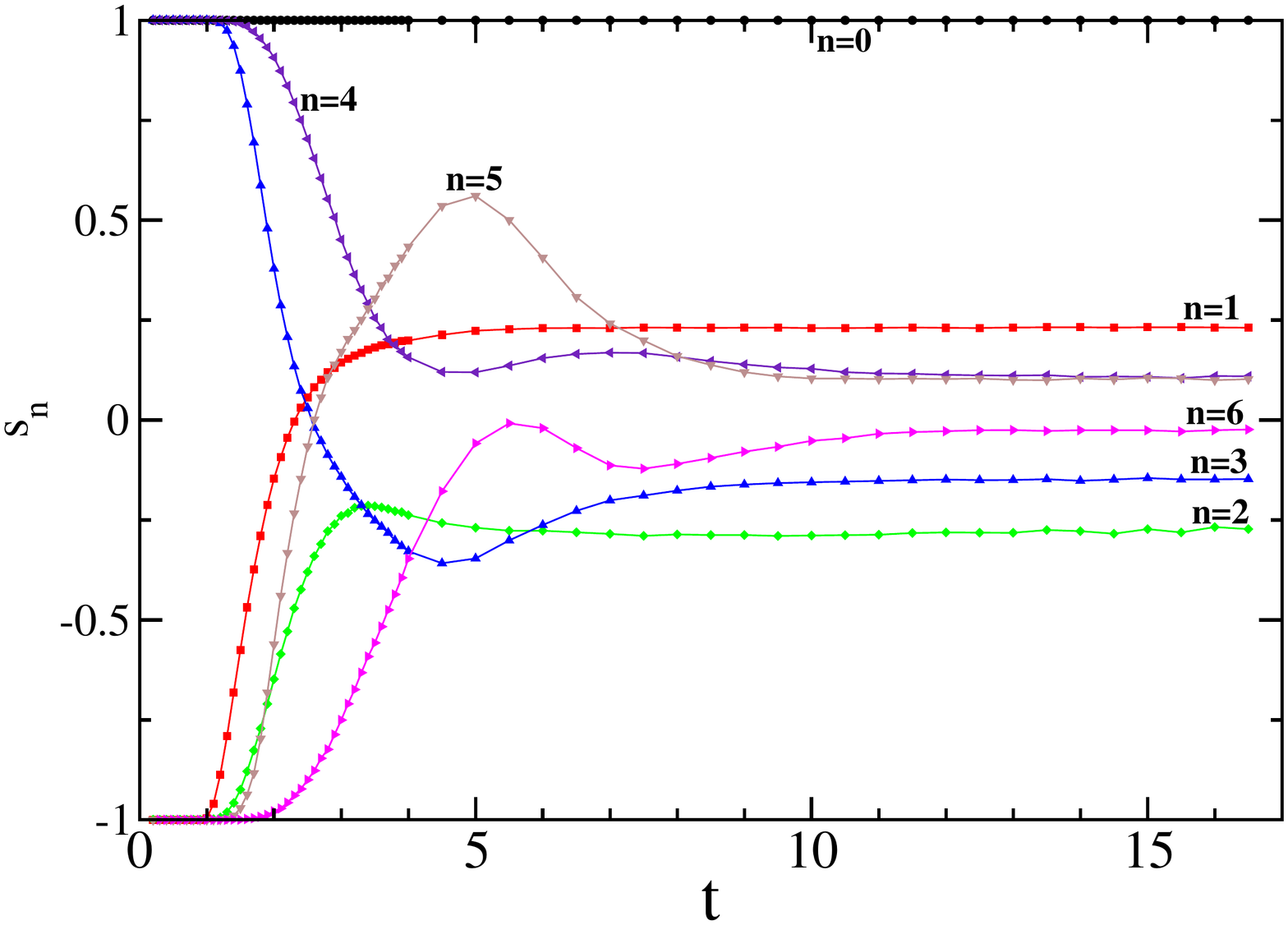}    
   \caption{\label{fig:sn} First 6 moments of the sign $s_n$ associated with the currents $I_n$ versus time $t$. Model B with $\alpha=0.5$, $\epsilon_d=0$, $\gamma=0.5$, $T=0$ and $V_b=0.5$. Assymptotic $s_6=0.025$.
   }
\end{figure}

\begin{figure}[h]
   \centering
    \includegraphics[width=9cm]{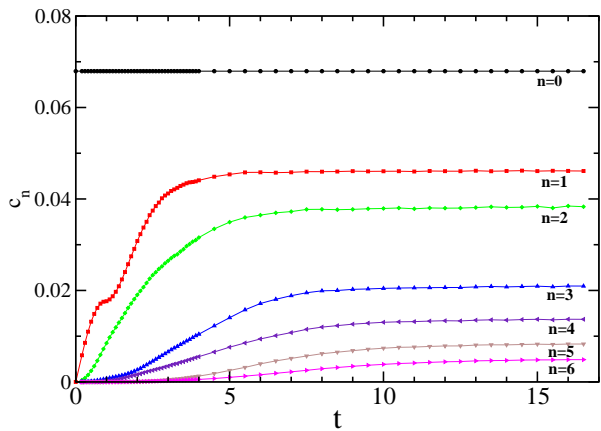}    
   \caption{\label{fig:cn} First 6 moments of the weight $c_n$ associated with the currents $I_n$ versus time $t$. Model B with $\alpha=0.5$, $\epsilon_d=0$, $\gamma=0.5$, $T=0$ and $V_b=0.5$.
   }
\end{figure}

Let us now analyze a bit further this calculation and present $I_n$,  the various orders of the expansion 
of the current in powers of $U$.  It is interesting to study how 
$I_n$ converge to the stationary value with time. Fig.~\ref{fig:In} shows the first seven moments 
$I_n$ (rescaled by $U^n$ with $U=3$ for visibility of the higher moments) as a function of time $t$.
At short time, the moments grow typically as $I_n \propto t^n$ which simply reflects the fact that $I_n$ is
an $n$-dimensional integral. After one or two oscillations, they reach their stationary value for roughly
$\Gamma t > 10$ but one notices that the higher orders converge significantly slower than the lower orders.
Simulations for large time are not particularly harder than for short time except for one small difficulty:
our insertion move has a flat distribution in the interval $[0,t]$ but only times close enough to $t$
actually contribute to the moments, so that the acceptance probability for these moves eventually drops when 
$t$ increases. This issue could be circonvaluated by reweighting the proposed moves around $t$. 
Fig.~\ref{fig:sn} and Fig.~\ref{fig:cn} show respectively the corresponding evolution of the signs $s_n$
and the weights $c_n$ as a function of time $t$. The signs remain rather large (the smallest value is $s_6=0.025$ asymptotically) and are not a limitation for the calculations. The decrease of the $I_n$ with $n$ essentially comes from the corresponding decrease of the $c_n$.
These Monte-Carlo computations are essentially free of the sign problem. The real limitation of the results presented in this section is a physical one: the apparent radius of convergence (in $U$) of the series appears to be finite, of the order of $1.5-2$ for the current.

\subsection{Magnetization}
\begin{figure}[h]
    \centering
    \includegraphics[width=8cm]{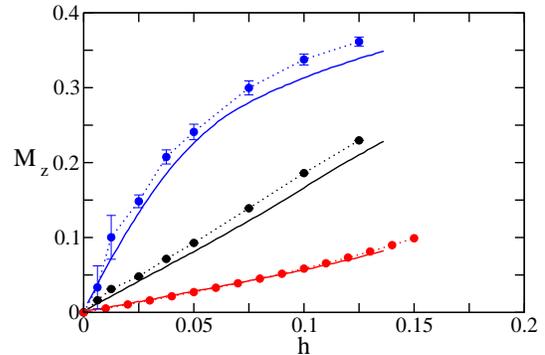}
    \caption{\label{fig:m-h} Zero temperature magnetization. $M_z=(Q_\uparrow-Q_\downarrow)/2$ as a function of magnetic field $h$ for model A with $\epsilon_d=-0.25$, $\gamma=1/4$, $T=0$ and $t=50$.
    Lines are analytical results (from Bethe-ansatz Ref.\onlinecite{okiji1982}), in particular red line is at $U=0$, black line at $U=0.7\Gamma$ ($\Gamma=4\gamma^2=1/4$), blue line at $U=1.5\Gamma$,
    dashed lines with points are results from QMC. The actual calculations were made with model B with $\alpha=0.5$ ($h\ge0.025$) and $\alpha=0.8$ ($h<0.025$),  while enforcing $\epsilon_d=\alpha U-0.25$.
    }
\end{figure}


In Fig.~\ref{fig:m-h}, we plot the magnetization $M_z=(Q_\uparrow-Q_\downarrow)/2$ as a function of the magnetic field $h$ for various strength of the interaction. At small field one expects
$M_z \propto h/T_K$ where the Kondo temperature $T_K$ is the characteristic scale of the Kondo effect. We do
observe indeed a sharp rise of the spin susceptibility when we switch on the electron-electron interaction.
As a consistency check, the full lines in  Fig.~\ref{fig:m-h} reproduce the results obtained in 
Ref.~\onlinecite{okiji1982} by solving the Bethe-ansatz equations. We find a remarkable agreement between the two (supposingly exact)
techniques. Note that the agreement is not supposed to be perfect as the Bethe-ansatz equations assume the
universal regime $\Gamma \ll D$ ($D=4$ is the band width) while the QMC calculations are performed for the microscopic model with a finite value $\Gamma=1/4$. Note that, as in the previous section, the results have been calculated at zero temperature and bias voltage which is the most difficult situation. Indeed,  at finite temperature the non-interacting Green's function decays faster with time, which ensures a faster convergence of the results with time and consequently smaller values of $c_n$. Following the strategy discussed in the previous section, the calculations are performed with a finite value of $\alpha$. 

\section{Discussion}


In summary, we have presented a general purpose algorithm to calculate systematically the electron-electron interaction corrections to the physical observables of a generic nanoelectronic circuit. We validate the approach using the non-equilibrium Anderson model and find that we can calculate the first moments (up to $n=15$ in this work using a recent laptop computer)  in the stationary regime and without being plagued by the appearance of the dynamical sign problem.
In a second step, one evaluates the interacting series whose radius of convergence is a priori unknown.
Our results indicate that this radius of convergence is strongly affected by the mean-field part of the interaction (role of the $\alpha$ parameter). 

There are many aspects left to future work. Clearly, one wants to apply the technique to other, larger 
systems such as a quantum point contact ("0.7" anomaly) or a quantum dot embedded in an interferometer. The 
technique could also be directly generalized to address several particle species (fermions or bosons), hence 
to study models coming from, e.g. circuit QED. Another route is to design techniques to obtain series with 
larger radius of convergence in order to reach more correlated regimes. This could be achieved using standard 
series analysis tools (Borel resummations...) but also by computing the perturbative series for other 
quantities, such as the self-energy. 
Last, one needs to develop the measurement of more complex objects, such as the Green's function, in order to be able to 
use the present technique for self-consistent schemes like DMFT or cluster DMFT. 

{\it Acknowledgments.}
E.P, C.G. X.W are funded by the ERC consolidator grant MesoQMC. 
L.M. and O.P. acknowledge support from the ERC consolidator grant MottMetals, under Grant Agreement No. 278472-MottMetals.
We thank S. Florens, M. Ferrero, E. Kozik, C. Mora, B. Nikolic and F. Werner for discussions and useful references.
We thank J.M. Luck for pointing out the Euler transform method and the associated references.

\appendix

\section{matrix $\nd{M}_n$ for density-density interaction}
In this Appendix, we study the form of the matrix $\nd{M}_n$ (defined in Eq.~\eqref{eq:M}) when the electron-electron interaction
takes the form of a density-density interaction:
\be
\hnd{H}_{\rm int}(t)=\sum_{ijkl} \nd{V}_{ij}(t)[\hnd{c}^\dagger_{i}\hnd{c}_i -\alpha_i][\hnd{c}^\dagger_{j}\hnd{c}_j-\alpha_j]
\ee
This form is more restrictive than the one studied in the main body of the manuscript. However, it allows for some optimizations as well as the inclusion of
the one-body correction (proportional to $\alpha_i$) which have proved to be important and is sufficient for the applications to a single interacting site shown in this article. The result is very simple: the matrix $\nd{M}_n$ must be replaced by $\nd{M}_n+\nd{L}_n$ where the diagonal matrix $\nd{L}_n$ consists of the $\alpha_i$:

\be
\nd{L}_n= -i \left(
\begin{array}{cccccc}
\alpha_{i_1} & 0            & 0           & 0           & \dots &0 \\
0            & \alpha_{j_1} & 0           & 0           & \dots &0 \\
0            & 0            &\alpha_{i_2} & 0           & \dots &0 \\
0            & 0            &0            &\alpha_{j_2} & \dots &0 \\
\dots        &\dots         &\dots        &\dots        & \dots &0 \\
0			 &0			   &0			 &0			   &\dots &0  \\
\end{array}
\right) 
\ee
This formula is from Ref.~\onlinecite{rubtsov2005}. We present here a simple recursive proof for completeness.
We need to evaluate averages of the form
\be
\Lambda =\langle (\hnd{c}^\dagger_1 \hnd{c}_1 -\alpha_1)(\hnd{c}^\dagger_2 \hnd{c}_2 -\alpha_2)\dots(\hnd{c}^\dagger_N \hnd{c}_N -\alpha_N) \rangle 
\ee
When all the $\alpha_i$ vanish, the above average is equal to $\Lambda ={\rm det}\ \nd{M}_n$. Let us now switch on $\alpha_1$. The previous result becomes $\Lambda ={\rm det}\ \nd{M}_n-\alpha_1 {\rm det}\ \nd{M}_n'$ where 
the  matrix $\nd{M}'_n$ is identical to $\nd{M}_n$ with
its first line and row removed. Equivalently, $\nd{M}'_n$ can be replaced by a matrix $\nd{R}_n$ of the same size as $\nd{M}_n$ with the first column filled with $(\alpha_1,0,\dots,0)$ and all the other columns are those of $\nd{M}_n$: one immediatly checks that upon developing the corresponding determinant with respect to the first column, one finds ${\rm det}\ \nd{R}_n=\alpha_1 {\rm det}\ \nd{M}'_n$. 
$\nd{R}_n$ and $\nd{M}_n$ having all their columns equal but the first one, they can now be put together into one  determinant of a single matrix $\Lambda ={\rm det}\ \nd{M}_n- {\rm det}\ \nd{R}_n = {\rm det}\ [\nd{M}_n +\nd{L}_n]$ (which is equal to $\nd{M}_n$ except for its upper corner shifted by $-\alpha_1$). 
This procedure can be continued with $\alpha_2,\alpha_3$\dots until no
$\alpha_i$ are left which proves the above statement. Note that in the particular case considered in this work, the matrix $\nd{M}_n+\nd{L}_n$ is block diagonal with respect to spin so that its determinant factorizes into two smaller determinants. This can be used for a faster calculation of the said determinants.

\section{Clustering property of the sum over Keldysh indices}
\label{app:clustering}

In this section, we show that summing over the Keldysh indices leads to a clustering property, 
{\it i.e.} that when the times of the integration in Eq.~\eqref{eq:basic} are far from the time $t$ where the observable is computed, 
the integrand decays. Let us consider Eq.~\eqref{eq:M}, in the case where e.g. $u_1,\dots, u_p$ are far from $t$, e.g. close to $t - \Delta t$ and 
$u_{p+1},\dots, u_n$ are close to $t$. We study the case where $\Delta t$ becomes large.
Let us denote $u_{p+1},\dots, u_n, t$ by $v_1, \dots, v_{n-p}$ to simplify the notations.
Then the matrix of  Eq.~\eqref{eq:M} has a block structure of the form:\
\be
\nd{M}_{n} =\left(
\begin{array}{ll}
A & B  \\
C & D 
\end{array}
\right)
\ee
where 
\begin{align}
  A_{i,j} &\equiv g(\bar u_i,\bar u_j)  \qquad  &B_{i,m} \equiv g(\bar u_i,\bar v_m)  \nonumber \\
  C_{l,j} &\equiv g(\bar v_l,\bar u_j)  \qquad  &D_{l,m} \equiv g(\bar v_l,\bar v_m) 
\end{align}
where $i,j=1, \dots,p$, and $l,m = 1, \dots n-p$.
Therefore 
\be 
\det \nd{M}_{n}({\cal C}_n,\{a_i\})  = \det A \det \bigl (D - B A^{-1} C \bigr )
\ee
Our assumption is that $|u_i - v_j| \sim O(\Delta t)$ for $\Delta t\rightarrow \infty$.
Moreover, at large time the non-interacting Green's function  decays, as $1/t$.
Since $B$ and $C$ contain only Green's functions with one $u$ and one $v$, their arguments are of order $\Delta t$,
and the matrix elements decay as $O(1/\Delta t)$.
Hence the matrix $\nd{M}_{n}$ is block-diagonal and 
\be 
\det \nd{M}_{n}({\cal C}_n,\{a_i\})  = \det A \det D  + O(1/\Delta t^2)
\ee
$A$ is in fact a $P_p$ matrix, which leads to 
\be
\sum_{a_1, \dots, a_p} (-1)^{\sum_{i=1}^p a_i} \det A =0
\ee
Moreover $D$ does not depend on the first $p$ Keldysh indices, which gives finally
\be 
\sum_{a_1, \dots, a_n}(-1)^{\sum_{i=1}^n a_i} \det \nd{M}_{n}({\cal C}_n,\{a_i\}) = O(1/\Delta t^2)
\ee
We see from this argument that at large separation (for $u_i$ far from $t$), the terms at fixed Keldysh indices
do not decay, while the sum over the Keldysh indices does. This is consistent with the observations made in 
Sect. \ref{brute}. This is also necessary for the coefficients $Q_n$ to converge at long time, {\it i.e.} for the 
long time limit and the $U$-expansion to commute.

\section{Properties of ${\bold\cal M}[{\cal C}_n]$}
\label{real}
\begin{widetext}
An important property of the $\nd{M}_{n}$ is found by considering the role of the {\it "mirror"} configurations
where the Keldysh indices $\{a_i \}$ are replaced by $\{1-a_i \}$.
Let us consider the local density first: $\hnd{c}^\dagger_i( t) \hnd{c}_i(t)$.
We suppose that the $u_i$ are ordered from the smallest to the largest (if not we relabel them). Then,
using Wick theorem, we have
\be
\det \nd{M}_{n}({\cal C}_n,\{a_i\}) = i^{2n+1}
\langle 
\hid{H}_{\rm int}( u_1)^{a_N} \hid{H}_{\rm int}( u_2)^{a_2}\dots \hid{H}_{\rm int}( u_n)^{a_1}
\hnd{c}^\dagger_i( t) \hnd{c}_i(t) 
\hid{H}_{\rm int}( u_n)^{1-a_N}\dots\hid{H}_{\rm int}( u_2)^{1-a_2}\hid{H}_{\rm int}( u_1)^{1-a_1}
\rangle
\ee
from which we immediately find that
\be
\det \nd{M}_{n}({\cal C}_n,\{a_i\})=- \det \nd{M}_{n}({\cal C}_n,\{1-a_i\})^*
\label{identity}
\ee
By adding together a configuration and its mirror,  we deduce from the above that
\be 
-i^{n+1} \sum_{\{a_i\}} (-1)^{\sum_i a_i} \det \nd{M}_{n}({\cal C}_n,\{a_i\}) 
\ee
is real and therefore ${\bold\cal M}[{\cal C}_n]$ from Eq.~\eqref{eq:yeh} is real as well.
The same argument holds for the current provided one replaces $\hnd{c}^\dagger_i( t) \hnd{c}_i(t) $ by
$i[ \hnd{c}^\dagger_i( t) \hnd{c}_j(t) -\hnd{c}^\dagger_j( t) \hnd{c}_i(t)].$ 
\end{widetext}

\section{Gray code for fast updates of the determinants}
\label{app:Gray}

The sum over the Keldysh indices can be accelerated using a Gray code, see e.g. Ref.~\onlinecite{numericalrecipes} section 20.2.
Using a Gray code, it is possible to enumerate all the integers from $0$ to $2^N -1$ in such a way that their binary representation change only by one bit
at each step, and that the last one contains only one bit (hence is also one bit flip away from 0).
To do this, we start with 0, and for $n=0,.. 2^N-1$ we flip the bit at position $c$ where $c$ is given by:
\begin{equation}
  c \equiv 
  \begin{cases}  
    \textrm{ffs}( \sim  n) & \text{ iif } n < 2^N-1 \\ 
    N & \text{ iif } n = 2^N-1 
  \end{cases}
\end{equation} 
where the $\textrm{ffs}(i)$ returns the position of the first (least significant) bit set in the integer $i$, 
and $\sim n$ is the binary complement of the integer $n$ (using the syntax of the {\it C} language).
We obtain all integers between $0$ and $2^N -1$ once and only once. The last case ensures that the integer returns to 0.

The Keldysh indices at order $N$ are a list of size $N$ of 0 (upper contour) and 1 (lower contour).
They are therefore in one-to-one correspondence with the integers from $0$ to $2^N -1$, via their binary representation.
Using the Gray code, we can enumerate the Keldysh indices by changing only one bit at a time, which in our algorithm means
changing one line and one column in the determinant at a time. This operation is of complexity $N^2$, 
i.e. much quicker than a full recomputation of the determinant $N^3$, 
and it is very commonly used in all determinantal Quantum Monte-Carlo. It is implemented using BLAS 2 operations.

The following piece of C++ code illustrates the use of the Gray code in our implementation:
\begin{lstlisting}
  auto two_to_N = uint64_t(1) << N; 
  for (uint64_t n = 0; n < two_to_N; ++n) { 
   int nlc = (n < two_to_N - 1 ? ffs(~n) : N); 
   // Change the line and column numbered nlc
   // in the matrix
  }
\end{lstlisting}
where $N$ is the the order at which we compute the determinant, 
Note that the matrix has returned after the loop to the value it had before the loop, up to numerical errors,
which are typically controlled at this place.

\section{Extrapolating series beyond their convergence radius}
\label{app:lindelof}
\begin{figure}[hbt]
    \centering
    \includegraphics[width=8cm]{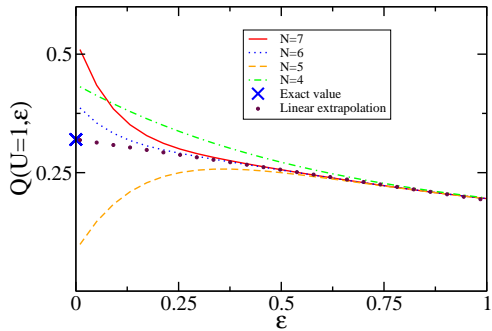}
    \caption{\label{fig:LindelofExtraTest} Extrapolation of the series. $Q(U=1,\epsilon)$ as a function of $\epsilon$ for model A with $\epsilon_d=0$, $T=0$ and $\gamma=0.5$ for various maximum order $N=4,5,6,7$. The circles show a simple extrapolation from a linear regression in the region where the various curves overlap. We do recover $Q(U=1)\approx 0.32$ well beyond the apparent convergence radius of the series.
    }
\end{figure}

In this appendix, we use the Lindel\"of\cite{hardy} extrapolation method, for the computation of the charge $Q(U)$ 
with $\alpha=0$. On Fig.~\ref{fig:modelA_2}, we have seen that for $\alpha=0$, the series can not be summed for $U>0.6$ directly.
Using the Lindel\"of formula: 
\be 
Q(U,N,\epsilon)\equiv \sum_{n=0}^N Q_n U^n e^{-\epsilon n \ln (n)}
\ee
we plot $Q(U,N,\epsilon)$ for $U=1.5$ as a function of $\epsilon$ in Fig.~\ref{fig:LindelofExtraTest}.
We see that the extrapolation method produces a result for $Q(U=1)$ in very good agreement with the imaginary-time QMC (dashed line in Fig.~\ref{fig:modelA_2}) result.

\section{Role of bound states in model A}
\begin{figure}[]
    \centering
    \includegraphics[width=8cm]{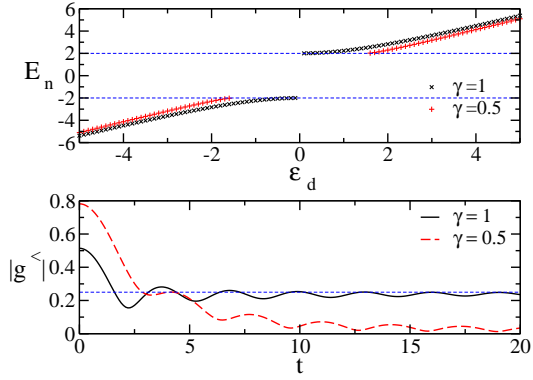}
    \caption{\label{fig:bound_state} Top panel: energy of the (unique) bound-state as a function of $\epsilon_d$. 
    Bottom panel: Absolute value of the diagonal part of the lesser Green's function $g_{00}^<(t)$ as a function of $t$, for $\xi_d=-0.5$ and $T=0$. 
    For $\gamma=0.5$, there is no bound state.
    }
\end{figure}

\begin{figure}[]
    \centering
    \includegraphics[width=8cm]{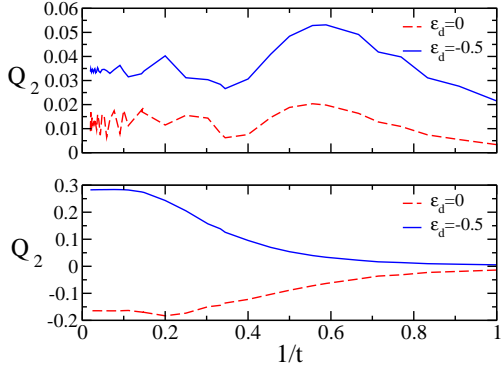}
    \caption{\label{fig:comp_1overt} $Q_2$ as a function of $1/t$, for $T=0$. 
    Top panel is for $\gamma=1$ (bound state), bottom panel for $\gamma=0.5$ (no bound state). 
    In each panel, the continuous (blue) line is for $\epsilon_d=0$, the dashed (red) line for $\epsilon_d=-0.5$
    }
\end{figure}

\begin{figure}[]
    \centering
    \includegraphics[width=8cm]{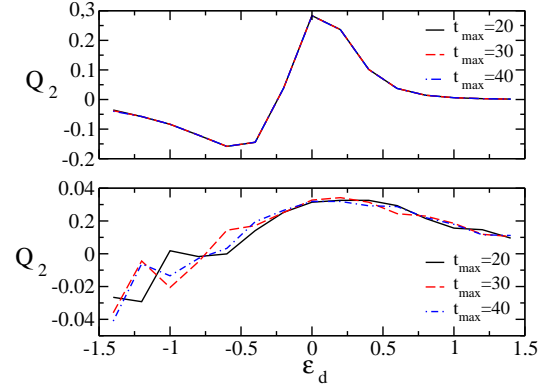}
    \caption{\label{fig:Q2vsE} $Q_2$ as a function of $\epsilon_d$, for $T=0$ and different values of $t$. 
    Top panel is for $\gamma=0.5$ (no bound state for $|\epsilon|<1.5$), bottom panel for $\gamma=1$ (bound state whatever $\epsilon_d$).
    dash-dotted line for $t=40$.
    }
\end{figure}

As an illustration of the role of bound states, we briefly study their role in model A. For this model, they can be obtained analytically by simple wave matching: One gets
\be
E_1=\frac{1}{2\gamma^2 -1} \left[ (\gamma^2-1)|\epsilon_d| +\gamma^2\sqrt{  \epsilon_d^2 + 4(2\gamma^2 -1)}\right]
\ee  
which exists when $|\epsilon_d|>2(1-\gamma^2)$. The top panel of Fig.~\ref{fig:bound_state} shows the corresponding bound state energy calculated numerically (which matches perfectly the above expression) as a function of $\epsilon_d$. When $\gamma < 1$, there is a finite window of values of $\epsilon_d$ where there are no bound states in the system. The bottom panel of Fig.~\ref{fig:bound_state} shows the associated non-interacting lesser Green's function of the system as a function of time. In the absence of bound state (dashed line, $\gamma=0.5$), we see that $g_{00}^<(t)$ decays towards zero at large time. Note that this decay would be much faster a larger temperature. In the presence of bound states however (full line, $\gamma=1$), we find that at large time $g_{00}^<$ saturates to its bound state contribution
$g_{00}^<(t)\propto e^{-iE_1 t}$. As a results, the convergence of the terms in the perturbative expansion with the time $t$ will be much slower and are not even guaranteed to converge as the bound states do not relax. This is illustrated in Fig.~\ref{fig:comp_1overt} which shows $Q_2$  (defined in Eq.~\eqref{eq:Qserie}) versus $1/t$ for $\gamma=1$ (upper panel, a bound state is present for any $\epsilon_d\ne 0$) and $\gamma=0.5$ (lower panel, no bound state). Similarly, Fig.~\ref{fig:Q2vsE} shows $Q_2$ versus $\epsilon_d$ with (lower panel) and without (upper panel) bound state. We find that without bound state, $Q_2$ quickly converges towards its stationary value (roughly for $t>10\Gamma$). However, in presence of bound states the convergence is much slower. 
\bibliographystyle{apsrev}
\bibliography{qmc_refs.bib}

\begin{thebibliography}{40}
\expandafter\ifx\csname natexlab\endcsname\relax\def\natexlab#1{#1}\fi
\expandafter\ifx\csname bibnamefont\endcsname\relax
  \def\bibnamefont#1{#1}\fi
\expandafter\ifx\csname bibfnamefont\endcsname\relax
  \def\bibfnamefont#1{#1}\fi
\expandafter\ifx\csname citenamefont\endcsname\relax
  \def\citenamefont#1{#1}\fi
\expandafter\ifx\csname url\endcsname\relax
  \def\url#1{\texttt{#1}}\fi
\expandafter\ifx\csname urlprefix\endcsname\relax\def\urlprefix{URL }\fi
\providecommand{\bibinfo}[2]{#2}
\providecommand{\eprint}[2][]{\url{#2}}


\bibitem[{\citenamefont{Lee and Ramakrishnan}(1985)}]{lee1985}
\bibinfo{author}{\bibfnamefont{P.~A.} \bibnamefont{Lee}} \bibnamefont{and}
  \bibinfo{author}{\bibfnamefont{T.~V.} \bibnamefont{Ramakrishnan}},
  \bibinfo{journal}{Rev. Mod. Phys.} \textbf{\bibinfo{volume}{57}},
  \bibinfo{pages}{287} (\bibinfo{year}{1985}),
  \urlprefix\url{http://link.aps.org/doi/10.1103/RevModPhys.57.287}.

\bibitem[{\citenamefont{Matveev and Larkin}(1992)}]{matveev1992}
\bibinfo{author}{\bibfnamefont{K.~A.} \bibnamefont{Matveev}} \bibnamefont{and}
  \bibinfo{author}{\bibfnamefont{A.~I.} \bibnamefont{Larkin}},
  \bibinfo{journal}{Phys. Rev. B} \textbf{\bibinfo{volume}{46}},
  \bibinfo{pages}{15337} (\bibinfo{year}{1992}),
  \urlprefix\url{http://link.aps.org/doi/10.1103/PhysRevB.46.15337}.

\bibitem[{\citenamefont{Takada et~al.}(2014)\citenamefont{Takada, B\"auerle,
  Yamamoto, Watanabe, Hermelin, Meunier, Alex, Weichselbaum, von Delft, Ludwig
  et~al.}}]{takada2014}
\bibinfo{author}{\bibfnamefont{S.}~\bibnamefont{Takada}},
  \bibinfo{author}{\bibfnamefont{C.}~\bibnamefont{B\"auerle}},
  \bibinfo{author}{\bibfnamefont{M.}~\bibnamefont{Yamamoto}},
  \bibinfo{author}{\bibfnamefont{K.}~\bibnamefont{Watanabe}},
  \bibinfo{author}{\bibfnamefont{S.}~\bibnamefont{Hermelin}},
  \bibinfo{author}{\bibfnamefont{T.}~\bibnamefont{Meunier}},
  \bibinfo{author}{\bibfnamefont{A.}~\bibnamefont{Alex}},
  \bibinfo{author}{\bibfnamefont{A.}~\bibnamefont{Weichselbaum}},
  \bibinfo{author}{\bibfnamefont{J.}~\bibnamefont{von Delft}},
  \bibinfo{author}{\bibfnamefont{A.}~\bibnamefont{Ludwig}},
  \bibnamefont{et~al.}, \bibinfo{journal}{Phys. Rev. Lett.}
  \textbf{\bibinfo{volume}{113}}, \bibinfo{pages}{126601}
  (\bibinfo{year}{2014}),
  \urlprefix\url{http://link.aps.org/doi/10.1103/PhysRevLett.113.126601}.

\bibitem[{\citenamefont{Thomas et~al.}(1996)\citenamefont{Thomas, Nicholls,
  Simmons, Pepper, Mace, and Ritchie}}]{thomas1996}
\bibinfo{author}{\bibfnamefont{K.~J.} \bibnamefont{Thomas}},
  \bibinfo{author}{\bibfnamefont{J.~T.} \bibnamefont{Nicholls}},
  \bibinfo{author}{\bibfnamefont{M.~Y.} \bibnamefont{Simmons}},
  \bibinfo{author}{\bibfnamefont{M.}~\bibnamefont{Pepper}},
  \bibinfo{author}{\bibfnamefont{D.~R.} \bibnamefont{Mace}}, \bibnamefont{and}
  \bibinfo{author}{\bibfnamefont{D.~A.} \bibnamefont{Ritchie}},
  \bibinfo{journal}{Phys. Rev. Lett.} \textbf{\bibinfo{volume}{77}},
  \bibinfo{pages}{135} (\bibinfo{year}{1996}),
  \urlprefix\url{http://link.aps.org/doi/10.1103/PhysRevLett.77.135}.

\bibitem[{\citenamefont{Schollw\"ock}(2005)}]{schollwock2005}
\bibinfo{author}{\bibfnamefont{U.}~\bibnamefont{Schollw\"ock}},
  \bibinfo{journal}{Rev. Mod. Phys.} \textbf{\bibinfo{volume}{77}},
  \bibinfo{pages}{259} (\bibinfo{year}{2005}),
  \urlprefix\url{http://link.aps.org/doi/10.1103/RevModPhys.77.259}.

\bibitem[{\citenamefont{Bulla et~al.}(2008)\citenamefont{Bulla, Costi, and
  Pruschke}}]{bulla2008}
\bibinfo{author}{\bibfnamefont{R.}~\bibnamefont{Bulla}},
  \bibinfo{author}{\bibfnamefont{T.~A.} \bibnamefont{Costi}}, \bibnamefont{and}
  \bibinfo{author}{\bibfnamefont{T.}~\bibnamefont{Pruschke}},
  \bibinfo{journal}{Rev. Mod. Phys.} \textbf{\bibinfo{volume}{80}},
  \bibinfo{pages}{395} (\bibinfo{year}{2008}),
  \urlprefix\url{http://link.aps.org/doi/10.1103/RevModPhys.80.395}.

\bibitem[{\citenamefont{{Gukelberger} et~al.}(2015)\citenamefont{{Gukelberger},
  {Huang}, and {Werner}}}]{gukelberg2015}
\bibinfo{author}{\bibfnamefont{J.}~\bibnamefont{{Gukelberger}}},
  \bibinfo{author}{\bibfnamefont{L.}~\bibnamefont{{Huang}}}, \bibnamefont{and}
  \bibinfo{author}{\bibfnamefont{P.}~\bibnamefont{{Werner}}},
  \bibinfo{journal}{ArXiv e-prints}  (\bibinfo{year}{2015}),
  \eprint{1501.04960}.

\bibitem[{\citenamefont{Prokof'ev and Svistunov}(2008)}]{Prokofiev08polaron}
\bibinfo{author}{\bibfnamefont{N.}~\bibnamefont{Prokof'ev}} \bibnamefont{and}
  \bibinfo{author}{\bibfnamefont{B.}~\bibnamefont{Svistunov}},
  \bibinfo{journal}{Phys. Rev. B} \textbf{\bibinfo{volume}{77}},
  \bibinfo{pages}{020408} (\bibinfo{year}{2008}).

\bibitem[{\citenamefont{Van~Houcke et~al.}(2008)\citenamefont{Van~Houcke,
  Kozik, Prokof'ev, and Svistunov}}]{DiagQMC}
\bibinfo{author}{\bibfnamefont{K.}~\bibnamefont{Van~Houcke}},
  \bibinfo{author}{\bibfnamefont{E.}~\bibnamefont{Kozik}},
  \bibinfo{author}{\bibfnamefont{N.}~\bibnamefont{Prokof'ev}},
  \bibnamefont{and}
  \bibinfo{author}{\bibfnamefont{B.}~\bibnamefont{Svistunov}}, in
  \emph{\bibinfo{booktitle}{Computer Simulation Studies in Condensed Matter
  Physics XXI}}, edited by
  \bibinfo{editor}{\bibfnamefont{D.}~\bibnamefont{Landau}},
  \bibinfo{editor}{\bibfnamefont{S.}~\bibnamefont{Lewis}}, \bibnamefont{and}
  \bibinfo{editor}{\bibfnamefont{H.}~\bibnamefont{Schuttler}}
  (\bibinfo{publisher}{Springer Verlag, Heidelberg}, \bibinfo{address}{Berlin},
  \bibinfo{year}{2008}).

\bibitem[{\citenamefont{Kozik et~al.}(2010)\citenamefont{Kozik, Houcke, Gull,
  Pollet, Prokof'ev, Svistunov, and Troyer}}]{KozikEPL10}
\bibinfo{author}{\bibfnamefont{E.}~\bibnamefont{Kozik}},
  \bibinfo{author}{\bibfnamefont{K.~V.} \bibnamefont{Houcke}},
  \bibinfo{author}{\bibfnamefont{E.}~\bibnamefont{Gull}},
  \bibinfo{author}{\bibfnamefont{L.}~\bibnamefont{Pollet}},
  \bibinfo{author}{\bibfnamefont{N.}~\bibnamefont{Prokof'ev}},
  \bibinfo{author}{\bibfnamefont{B.}~\bibnamefont{Svistunov}},
  \bibnamefont{and} \bibinfo{author}{\bibfnamefont{M.}~\bibnamefont{Troyer}},
  \bibinfo{journal}{EPL (Europhysics Letters)} \textbf{\bibinfo{volume}{90}},
  \bibinfo{pages}{10004} (\bibinfo{year}{2010}),
  \urlprefix\url{http://stacks.iop.org/0295-5075/90/i=1/a=10004}.

\bibitem[{\citenamefont{Van~Houcke et~al.}(2012)\citenamefont{Van~Houcke,
  Werner, Kozik, Prokof’ev, Svistunov, Ku, Sommer, Cheuk, Schirotzek, and
  Zwierlein}}]{van2012NatPhys}
\bibinfo{author}{\bibfnamefont{K.}~\bibnamefont{Van~Houcke}},
  \bibinfo{author}{\bibfnamefont{F.}~\bibnamefont{Werner}},
  \bibinfo{author}{\bibfnamefont{E.}~\bibnamefont{Kozik}},
  \bibinfo{author}{\bibfnamefont{N.}~\bibnamefont{Prokof’ev}},
  \bibinfo{author}{\bibfnamefont{B.}~\bibnamefont{Svistunov}},
  \bibinfo{author}{\bibfnamefont{M.}~\bibnamefont{Ku}},
  \bibinfo{author}{\bibfnamefont{A.}~\bibnamefont{Sommer}},
  \bibinfo{author}{\bibfnamefont{L.}~\bibnamefont{Cheuk}},
  \bibinfo{author}{\bibfnamefont{A.}~\bibnamefont{Schirotzek}},
  \bibnamefont{and}
  \bibinfo{author}{\bibfnamefont{M.}~\bibnamefont{Zwierlein}},
  \bibinfo{journal}{Nature Physics} \textbf{\bibinfo{volume}{8}},
  \bibinfo{pages}{366} (\bibinfo{year}{2012}).

\bibitem[{\citenamefont{Rubtsov et~al.}(2005)\citenamefont{Rubtsov, Savkin, and
  Lichtenstein}}]{rubtsov2005}
\bibinfo{author}{\bibfnamefont{A.~N.} \bibnamefont{Rubtsov}},
  \bibinfo{author}{\bibfnamefont{V.~V.} \bibnamefont{Savkin}},
  \bibnamefont{and} \bibinfo{author}{\bibfnamefont{A.~I.}
  \bibnamefont{Lichtenstein}}, \bibinfo{journal}{Phys. Rev. B}
  \textbf{\bibinfo{volume}{72}}, \bibinfo{pages}{035122}
  (\bibinfo{year}{2005}),
  \urlprefix\url{http://link.aps.org/doi/10.1103/PhysRevB.72.035122}.

\bibitem[{\citenamefont{Werner et~al.}(2006)\citenamefont{Werner, Comanac, de'
  Medici, Troyer, and Millis}}]{werner2006}
\bibinfo{author}{\bibfnamefont{P.}~\bibnamefont{Werner}},
  \bibinfo{author}{\bibfnamefont{A.}~\bibnamefont{Comanac}},
  \bibinfo{author}{\bibfnamefont{L.}~\bibnamefont{de' Medici}},
  \bibinfo{author}{\bibfnamefont{M.}~\bibnamefont{Troyer}}, \bibnamefont{and}
  \bibinfo{author}{\bibfnamefont{A.~J.} \bibnamefont{Millis}},
  \bibinfo{journal}{Phys. Rev. Lett.} \textbf{\bibinfo{volume}{97}},
  \bibinfo{pages}{076405} (\bibinfo{year}{2006}),
  \urlprefix\url{http://link.aps.org/doi/10.1103/PhysRevLett.97.076405}.

\bibitem[{\citenamefont{Gull et~al.}(2008)\citenamefont{Gull, Werner,
  Parcollet, and Troyer}}]{gull2008}
\bibinfo{author}{\bibfnamefont{E.}~\bibnamefont{Gull}},
  \bibinfo{author}{\bibfnamefont{P.}~\bibnamefont{Werner}},
  \bibinfo{author}{\bibfnamefont{O.}~\bibnamefont{Parcollet}},
  \bibnamefont{and} \bibinfo{author}{\bibfnamefont{M.}~\bibnamefont{Troyer}},
  \bibinfo{journal}{EPL (Europhysics Letters)} \textbf{\bibinfo{volume}{82}},
  \bibinfo{pages}{57003} (\bibinfo{year}{2008}),
  \urlprefix\url{http://stacks.iop.org/0295-5075/82/i=5/a=57003}.

\bibitem[{\citenamefont{Gull et~al.}(2011)\citenamefont{Gull, Millis,
  Lichtenstein, Rubtsov, Troyer, and Werner}}]{gull2011}
\bibinfo{author}{\bibfnamefont{E.}~\bibnamefont{Gull}},
  \bibinfo{author}{\bibfnamefont{A.~J.} \bibnamefont{Millis}},
  \bibinfo{author}{\bibfnamefont{A.~I.} \bibnamefont{Lichtenstein}},
  \bibinfo{author}{\bibfnamefont{A.~N.} \bibnamefont{Rubtsov}},
  \bibinfo{author}{\bibfnamefont{M.}~\bibnamefont{Troyer}}, \bibnamefont{and}
  \bibinfo{author}{\bibfnamefont{P.}~\bibnamefont{Werner}},
  \bibinfo{journal}{Rev. Mod. Phys.} \textbf{\bibinfo{volume}{83}},
  \bibinfo{pages}{349} (\bibinfo{year}{2011}),
  \urlprefix\url{http://link.aps.org/doi/10.1103/RevModPhys.83.349}.

\bibitem[{\citenamefont{M\"uhlbacher and Rabani}(2008)}]{muhlbacher2008}
\bibinfo{author}{\bibfnamefont{L.}~\bibnamefont{M\"uhlbacher}}
  \bibnamefont{and} \bibinfo{author}{\bibfnamefont{E.}~\bibnamefont{Rabani}},
  \bibinfo{journal}{Phys. Rev. Lett.} \textbf{\bibinfo{volume}{100}},
  \bibinfo{pages}{176403} (\bibinfo{year}{2008}),
  \urlprefix\url{http://link.aps.org/doi/10.1103/PhysRevLett.100.176403}.

\bibitem[{\citenamefont{Schir\'o and Fabrizio}(2009)}]{schiro2009}
\bibinfo{author}{\bibfnamefont{M.}~\bibnamefont{Schir\'o}} \bibnamefont{and}
  \bibinfo{author}{\bibfnamefont{M.}~\bibnamefont{Fabrizio}},
  \bibinfo{journal}{Phys. Rev. B} \textbf{\bibinfo{volume}{79}},
  \bibinfo{pages}{153302} (\bibinfo{year}{2009}),
  \urlprefix\url{http://link.aps.org/doi/10.1103/PhysRevB.79.153302}.

\bibitem[{\citenamefont{Schir\'o}(2010)}]{schiro2010}
\bibinfo{author}{\bibfnamefont{M.}~\bibnamefont{Schir\'o}},
  \bibinfo{journal}{Phys. Rev. B} \textbf{\bibinfo{volume}{81}},
  \bibinfo{pages}{085126} (\bibinfo{year}{2010}),
  \urlprefix\url{http://link.aps.org/doi/10.1103/PhysRevB.81.085126}.

\bibitem[{\citenamefont{Werner et~al.}(2009)\citenamefont{Werner, Oka, and
  Millis}}]{werner2009}
\bibinfo{author}{\bibfnamefont{P.}~\bibnamefont{Werner}},
  \bibinfo{author}{\bibfnamefont{T.}~\bibnamefont{Oka}}, \bibnamefont{and}
  \bibinfo{author}{\bibfnamefont{A.~J.} \bibnamefont{Millis}},
  \bibinfo{journal}{Phys. Rev. B} \textbf{\bibinfo{volume}{79}},
  \bibinfo{pages}{035320} (\bibinfo{year}{2009}),
  \urlprefix\url{http://link.aps.org/doi/10.1103/PhysRevB.79.035320}.

\bibitem[{\citenamefont{Werner et~al.}(2010)\citenamefont{Werner, Oka,
  Eckstein, and Millis}}]{werner2010}
\bibinfo{author}{\bibfnamefont{P.}~\bibnamefont{Werner}},
  \bibinfo{author}{\bibfnamefont{T.}~\bibnamefont{Oka}},
  \bibinfo{author}{\bibfnamefont{M.}~\bibnamefont{Eckstein}}, \bibnamefont{and}
  \bibinfo{author}{\bibfnamefont{A.~J.} \bibnamefont{Millis}},
  \bibinfo{journal}{Phys. Rev. B} \textbf{\bibinfo{volume}{81}},
  \bibinfo{pages}{035108} (\bibinfo{year}{2010}),
  \urlprefix\url{http://link.aps.org/doi/10.1103/PhysRevB.81.035108}.

\bibitem[{\citenamefont{Cohen et~al.}(2014{\natexlab{a}})\citenamefont{Cohen,
  Gull, Reichman, and Millis}}]{cohen2014greenPRL}
\bibinfo{author}{\bibfnamefont{G.}~\bibnamefont{Cohen}},
  \bibinfo{author}{\bibfnamefont{E.}~\bibnamefont{Gull}},
  \bibinfo{author}{\bibfnamefont{D.~R.} \bibnamefont{Reichman}},
  \bibnamefont{and} \bibinfo{author}{\bibfnamefont{A.~J.}
  \bibnamefont{Millis}}, \bibinfo{journal}{Physical review letters}
  \textbf{\bibinfo{volume}{112}}, \bibinfo{pages}{146802}
  (\bibinfo{year}{2014}{\natexlab{a}}).

\bibitem[{\citenamefont{Cohen et~al.}(2014{\natexlab{b}})\citenamefont{Cohen,
  Reichman, Millis, and Gull}}]{cohen2014greenPRB}
\bibinfo{author}{\bibfnamefont{G.}~\bibnamefont{Cohen}},
  \bibinfo{author}{\bibfnamefont{D.~R.} \bibnamefont{Reichman}},
  \bibinfo{author}{\bibfnamefont{A.~J.} \bibnamefont{Millis}},
  \bibnamefont{and} \bibinfo{author}{\bibfnamefont{E.}~\bibnamefont{Gull}},
  \bibinfo{journal}{Physical Review B} \textbf{\bibinfo{volume}{89}},
  \bibinfo{pages}{115139} (\bibinfo{year}{2014}{\natexlab{b}}).

\bibitem[{\citenamefont{Groth et~al.}(2014)\citenamefont{Groth, Wimmer,
  Akhmerov, and Waintal}}]{groth2014}
\bibinfo{author}{\bibfnamefont{C.~W.} \bibnamefont{Groth}},
  \bibinfo{author}{\bibfnamefont{M.}~\bibnamefont{Wimmer}},
  \bibinfo{author}{\bibfnamefont{A.~R.} \bibnamefont{Akhmerov}},
  \bibnamefont{and} \bibinfo{author}{\bibfnamefont{X.}~\bibnamefont{Waintal}},
  \bibinfo{journal}{New J. Phys.} \textbf{\bibinfo{volume}{16}},
  \bibinfo{pages}{063065} (\bibinfo{year}{2014}).

\bibitem[{\citenamefont{Parcollet et~al.}(2014)\citenamefont{Parcollet,
  Ferrero, Ayral, Hafermann, Seth, and Krivenko}}]{TRIQS}
\bibinfo{author}{\bibfnamefont{O.}~\bibnamefont{Parcollet}},
  \bibinfo{author}{\bibfnamefont{M.}~\bibnamefont{Ferrero}},
  \bibinfo{author}{\bibfnamefont{T.}~\bibnamefont{Ayral}},
  \bibinfo{author}{\bibfnamefont{H.}~\bibnamefont{Hafermann}},
  \bibinfo{author}{\bibfnamefont{P.}~\bibnamefont{Seth}}, \bibnamefont{and}
  \bibinfo{author}{\bibfnamefont{I.~S.} \bibnamefont{Krivenko}},
  \bibinfo{journal}{in preparation}  (\bibinfo{year}{2014}).

\bibitem[{\citenamefont{Shi and Zhang}(2013)}]{Shi2013}
\bibinfo{author}{\bibfnamefont{H.}~\bibnamefont{Shi}} \bibnamefont{and}
  \bibinfo{author}{\bibfnamefont{S.}~\bibnamefont{Zhang}},
  \bibinfo{journal}{Phys. Rev. B} \textbf{\bibinfo{volume}{88}},
  \bibinfo{pages}{125132} (\bibinfo{year}{2013}),
  \urlprefix\url{http://link.aps.org/doi/10.1103/PhysRevB.88.125132}.

\bibitem[{\citenamefont{Rom et~al.}(1997)\citenamefont{Rom, Charutz1, and
  Neuhauser}}]{Naomi1997}
\bibinfo{author}{\bibfnamefont{N.}~\bibnamefont{Rom}},
  \bibinfo{author}{\bibfnamefont{D.}~\bibnamefont{Charutz1}}, \bibnamefont{and}
  \bibinfo{author}{\bibfnamefont{D.}~\bibnamefont{Neuhauser}},
  \bibinfo{journal}{Chemical Physics Letters} \textbf{\bibinfo{volume}{270}},
  \bibinfo{pages}{382} (\bibinfo{year}{1997}).

\bibitem[{\citenamefont{Meir and Wingreen}(1992)}]{meir1992}
\bibinfo{author}{\bibfnamefont{Y.}~\bibnamefont{Meir}} \bibnamefont{and}
  \bibinfo{author}{\bibfnamefont{N.~S.} \bibnamefont{Wingreen}},
  \bibinfo{journal}{Phys. Rev. Lett.} \textbf{\bibinfo{volume}{68}},
  \bibinfo{pages}{2512} (\bibinfo{year}{1992}),
  \urlprefix\url{http://link.aps.org/doi/10.1103/PhysRevLett.68.2512}.

\bibitem[{\citenamefont{Gaury et~al.}(2014)\citenamefont{Gaury, Weston, Santin,
  Houzet, Groth, and Waintal}}]{gaury2014}
\bibinfo{author}{\bibfnamefont{B.}~\bibnamefont{Gaury}},
  \bibinfo{author}{\bibfnamefont{J.}~\bibnamefont{Weston}},
  \bibinfo{author}{\bibfnamefont{M.}~\bibnamefont{Santin}},
  \bibinfo{author}{\bibfnamefont{M.}~\bibnamefont{Houzet}},
  \bibinfo{author}{\bibfnamefont{C.}~\bibnamefont{Groth}}, \bibnamefont{and}
  \bibinfo{author}{\bibfnamefont{X.}~\bibnamefont{Waintal}},
  \bibinfo{journal}{Physics Reports} \textbf{\bibinfo{volume}{534}},
  \bibinfo{pages}{1} (\bibinfo{year}{2014}).

\bibitem[{\citenamefont{Rammer and Smith}(1986)}]{rammer1986}
\bibinfo{author}{\bibfnamefont{J.}~\bibnamefont{Rammer}} \bibnamefont{and}
  \bibinfo{author}{\bibfnamefont{H.}~\bibnamefont{Smith}},
  \bibinfo{journal}{Rev. Mod. Phys.} \textbf{\bibinfo{volume}{58}},
  \bibinfo{pages}{323} (\bibinfo{year}{1986}),
  \urlprefix\url{http://link.aps.org/doi/10.1103/RevModPhys.58.323}.

\bibitem[{\citenamefont{Gaury and Waintal}(2014)}]{gaury2014b}
\bibinfo{author}{\bibfnamefont{B.}~\bibnamefont{Gaury}} \bibnamefont{and}
  \bibinfo{author}{\bibfnamefont{X.}~\bibnamefont{Waintal}},
  \bibinfo{journal}{Nat Commun} \textbf{\bibinfo{volume}{5}},
  \bibinfo{pages}{3844} (\bibinfo{year}{2014}).

\bibitem[{\citenamefont{Prokof'ev et~al.}(1996)\citenamefont{Prokof'ev,
  Svistunov, and Tupitsyn}}]{Prokofiev96}
\bibinfo{author}{\bibfnamefont{N.}~\bibnamefont{Prokof'ev}},
  \bibinfo{author}{\bibfnamefont{B.}~\bibnamefont{Svistunov}},
  \bibnamefont{and} \bibinfo{author}{\bibfnamefont{I.}~\bibnamefont{Tupitsyn}},
  \bibinfo{journal}{Pis’ma v Zh.Eks. Teor. Fiz.}
  \textbf{\bibinfo{volume}{64}}, \bibinfo{pages}{853} (\bibinfo{year}{1996}),
  \bibinfo{note}{english translation is Report cond-mat/9612091}.

\bibitem[{\citenamefont{H.}(1949)}]{hardy}
\bibinfo{author}{\bibfnamefont{H.~G.} \bibnamefont{H.}},
  \emph{\bibinfo{title}{Divergent Series}} (\bibinfo{publisher}{Edition Jacques
  Gabay}, \bibinfo{address}{New York, NY, USA}, \bibinfo{year}{1949}),
  \bibinfo{edition}{3rd} ed.

\bibitem[{\citenamefont{{Mera} et~al.}(2014)\citenamefont{{Mera}, {Pedersen},
  and {Nikolic}}}]{mera2014}
\bibinfo{author}{\bibfnamefont{H.}~\bibnamefont{{Mera}}},
  \bibinfo{author}{\bibfnamefont{T.~G.} \bibnamefont{{Pedersen}}},
  \bibnamefont{and} \bibinfo{author}{\bibfnamefont{B.~K.}
  \bibnamefont{{Nikolic}}}, \bibinfo{journal}{ArXiv e-prints}
  (\bibinfo{year}{2014}), \eprint{1405.7956}.

\bibitem[{\citenamefont{Guant and Guttmann}(1974)}]{DombGreen}
\bibinfo{author}{\bibfnamefont{A.}~\bibnamefont{Guant}} \bibnamefont{and}
  \bibinfo{author}{\bibfnamefont{D.}~\bibnamefont{Guttmann}},
  \emph{\bibinfo{title}{Asymptotic Analysis of Coefficients}}
  (\bibinfo{publisher}{Academic Press, New York}, \bibinfo{year}{1974}).

\bibitem[{\citenamefont{{Pustilnik} and {Glazman}}(2004)}]{pustilnik2004}
\bibinfo{author}{\bibfnamefont{M.}~\bibnamefont{{Pustilnik}}} \bibnamefont{and}
  \bibinfo{author}{\bibfnamefont{L.}~\bibnamefont{{Glazman}}},
  \bibinfo{journal}{Journal of Physics Condensed Matter}
  \textbf{\bibinfo{volume}{16}}, \bibinfo{pages}{513} (\bibinfo{year}{2004}),
  \eprint{cond-mat/0401517}.

\bibitem[{\citenamefont{{Mora} et~al.}(2014)\citenamefont{{Mora}, {Pascu Moca},
  {von Delft}, and {Zarand}}}]{mora2014}
\bibinfo{author}{\bibfnamefont{C.}~\bibnamefont{{Mora}}},
  \bibinfo{author}{\bibfnamefont{C.}~\bibnamefont{{Pascu Moca}}},
  \bibinfo{author}{\bibfnamefont{J.}~\bibnamefont{{von Delft}}},
  \bibnamefont{and} \bibinfo{author}{\bibfnamefont{G.}~\bibnamefont{{Zarand}}},
  \bibinfo{journal}{ArXiv e-prints}  (\bibinfo{year}{2014}),
  \eprint{1409.3451}.

\bibitem[{\citenamefont{Okiji and Kawakami}(1982)}]{okiji1982}
\bibinfo{author}{\bibfnamefont{A.}~\bibnamefont{Okiji}} \bibnamefont{and}
  \bibinfo{author}{\bibfnamefont{N.}~\bibnamefont{Kawakami}},
  \bibinfo{journal}{Journal of the Physical Society of Japan}
  \textbf{\bibinfo{volume}{51}}, \bibinfo{pages}{3192} (\bibinfo{year}{1982}).

\bibitem[{\citenamefont{Press et~al.}(1992)\citenamefont{Press, Teukolsky,
  Vetterling, and Flannery}}]{numericalrecipes}
\bibinfo{author}{\bibfnamefont{W.~H.} \bibnamefont{Press}},
  \bibinfo{author}{\bibfnamefont{S.~A.} \bibnamefont{Teukolsky}},
  \bibinfo{author}{\bibfnamefont{W.~T.} \bibnamefont{Vetterling}},
  \bibnamefont{and} \bibinfo{author}{\bibfnamefont{B.~P.}
  \bibnamefont{Flannery}}, \emph{\bibinfo{title}{Numerical Recipes in C: The
  Art of Scientific Computing. Second Edition}} (\bibinfo{publisher}{Cambridge
  University Press}, \bibinfo{address}{New York, NY, USA},
  \bibinfo{year}{1992}), \bibinfo{edition}{2nd} ed.

\end{thebibliography}

\end{document}